\documentclass[aip,cha,amsmath,amssymb,reprint]{revtex4-1}

\usepackage{graphicx}% Include figure files
\usepackage{dcolumn}% Align table columns on decimal point
\usepackage{bm}% bold math
\usepackage[utf8]{inputenc}
\usepackage[T1]{fontenc}
\usepackage{mathptmx}
\usepackage{etoolbox}
\usepackage{xcolor}

\newcommand{\iu}{{i\mkern1mu}}
\newcommand{\numrea}{\ensuremath{N_{\mathrm{r}}}}

\makeatletter
\def\@email#1#2{%
 \endgroup
 \patchcmd{\titleblock@produce}
  {\frontmatter@RRAPformat}
  {\frontmatter@RRAPformat{\produce@RRAP{*#1\href{mailto:#2}{#2}}}\frontmatter@RRAPformat}
  {}{}
}%
\makeatother

\begin{document}

\title{Synchronization dynamics of phase oscillators on power grid models}
\author{Max Potratzki}
\affiliation{Department of Epileptology, University of Bonn Medical Centre, Venusberg Campus 1, 53127 Bonn, Germany}
\author{Timo Br\"ohl}
\affiliation{Department of Epileptology, University of Bonn Medical Centre, Venusberg Campus 1, 53127 Bonn, Germany}
\affiliation{Helmholtz Institute for Radiation and Nuclear Physics, University of Bonn, Nussallee 14--16, 53115 Bonn, Germany}
\author{Thorsten Rings}
\affiliation{Department of Epileptology, University of Bonn Medical Centre, Venusberg Campus 1, 53127 Bonn, Germany}
\affiliation{Helmholtz Institute for Radiation and Nuclear Physics, University of Bonn, Nussallee 14--16, 53115 Bonn, Germany}
\author{Klaus Lehnertz}
\email{klaus.lehnertz@ukbonn.de}
\affiliation{Department of Epileptology, University of Bonn Medical Centre, Venusberg Campus 1, 53127 Bonn, Germany}
\affiliation{Helmholtz Institute for Radiation and Nuclear Physics, University of Bonn, Nussallee 14--16, 53115 Bonn, Germany}
\affiliation{Interdisciplinary Center for Complex Systems, University of Bonn, Br{\"u}hler Stra\ss{}e 7, 53175 Bonn, Germany}

\date{\today}% It is always \today, today,

\begin{abstract}
We investigate topological and spectral properties of models of European and US-American power grids and of paradigmatic network models as well as their implications for the synchronization dynamics of phase oscillators with heterogeneous natural frequencies.
We employ the complex-valued order parameter --~a widely-used indicator for phase ordering~-- to assess the synchronization dynamics and observe the order parameter to exhibit either constant or periodic or non-periodic, possibly chaotic temporal evolutions for a given coupling strength but depending on initial conditions and the systems' disorder.
Interestingly, both topological and spectral characteristics of the power grids point to a diminished capability of these networks to support a temporarily stable synchronization dynamics.
We find non-trivial commonalities between the synchronization dynamics of oscillators on seemingly opposing topologies.
\end{abstract}

\maketitle

\begin{quotation}
Many natural and man-made systems can be described as phase oscillators coupled onto a network with a complex interaction topology.
We here report nontrivial synchronization dynamics of simple phase oscillators with heterogeneous natural frequencies coupled onto power grid models. 
Although many of these dynamics resemble the ones seen for well-studied paradigmatic networks, relevant topological and spectral properties of the latter largely differ from those of the power grid models.  
These properties could be the reason for the diminished capability of the power grid models to support a stable synchronization dynamics.
\end{quotation}

%%%%%%%%%%%%%%%%%%%%%%%%%%%%%%%%%%%%%%%%%%%%%%%%%%%%%%%%%%%%%%%%%%%%%%%%%%
\section{Introduction}
%%%%%%%%%%%%%%%%%%%%%%%%%%%%%%%%%%%%%%%%%%%%%%%%%%%%%%%%%%%%%%%%%%%%%%%%%%
%
Synchronization and related complex phenomena in populations of interacting elements are ubiquitous in nature and play an important role in numerous scientific fields, ranging from physics to the neurosciences~\cite{pikovsky2001,glass2001,boccaletti2002,mosekilde2002,Rosenblum2003,Bennett2004,fell-axmacher2011,Pikovsky2015,Boccaletti2018}, and in technology~\cite{blekhman1988,Nijmeijer2003,kapitaniak2012,Motter2013,Doerfler2014,Csaba2020,Witthaut2022}.
In networks of interacting elements, synchronization emerges from the complex interplay between network topology and vertex dynamics~\cite{gomez2007,gomez2007b,arenas2008,Rohden2014}.  
An improved understanding of the various forms of synchronization can be achieved by modeling each element of the population as an oscillator.
Among the many models available, the Kuramoto model is often used in various contexts~\cite{strogatz2000,acebron2005,Breakspear2010,rodrigues2016,Bick2020}.
It consists of a population of $N$ globally coupled phase oscillators, and the population's macroscopic state can be characterized by the complex-valued order parameter~\cite{kuramoto1984,Schroeder2017}.
The long-time average~\cite{Ott2009,Mirollo2012} of its absolute value is usually used as a single measure for phase ordering. 
There are, however, systems for which the temporal evolution of the order parameter exhibits large fluctuations or might even indicate a chaotic motion~\cite{Bick2018,smith2019, clusella2020}.
These include systems with delays~\cite{Yeung1999,Senthilkumar2011,Gjurchinovski2017}, systems with nontrivial coupling topologies~\cite{Bick2011,Labavic2017,Chouzouris2018,Paolini2022,Tadic2022}, and time-dependent oscillator networks~\cite{Buscarino2015,Assenza2011,Ghosh2022}.
Nontrivial temporal evolutions of the order parameter were also reported for other oscillator models~\cite{rothkegel2011,rothkegel2014,Gerster2020,Boaretto2021,LeeKrischner2021,Clusella2022,Sawicki2022,Thiele2023}.

Here, we investigate synchronization dynamics of Kuramoto phase oscillators coupled onto frequently-used models of power grids and paradigmatic network models.
Considering heterogeneities seen in real power grids, we find the oscillators' synchronization dynamics to vary vastly depending on initial conditions  and the systems' disorder.
We conjecture that the variabilities can be traced back, at least in part, to topological and spectral properties of the networks.

%%%%%%%%%%%%%%%%%%%%%%%%%%%%%%%%%%%%%%%%%%%%%%%%%%%%%%%%%%%%%%%%%%%%%%%%%%
\section{Methods}
%%%%%%%%%%%%%%%%%%%%%%%%%%%%%%%%%%%%%%%%%%%%%%%%%%%%%%%%%%%%%%%%%%%%%%%%%%

%%%%%%%%%%%%%%%%%%%%%%%%%%%%%%%%%%%%%%%%%%%%%%%%%%%%%%%%%%%%%%%%%%%%%%%%%%
\subsection{Power grid models}
%%%%%%%%%%%%%%%%%%%%%%%%%%%%%%%%%%%%%%%%%%%%%%%%%%%%%%%%%%%%%%%%%%%%%%%%%%
%
For our investigations, we make use of two publicly available power grid datasets.
\begin{figure}[htbp]
	\centering
	\includegraphics[width=0.45\textwidth]{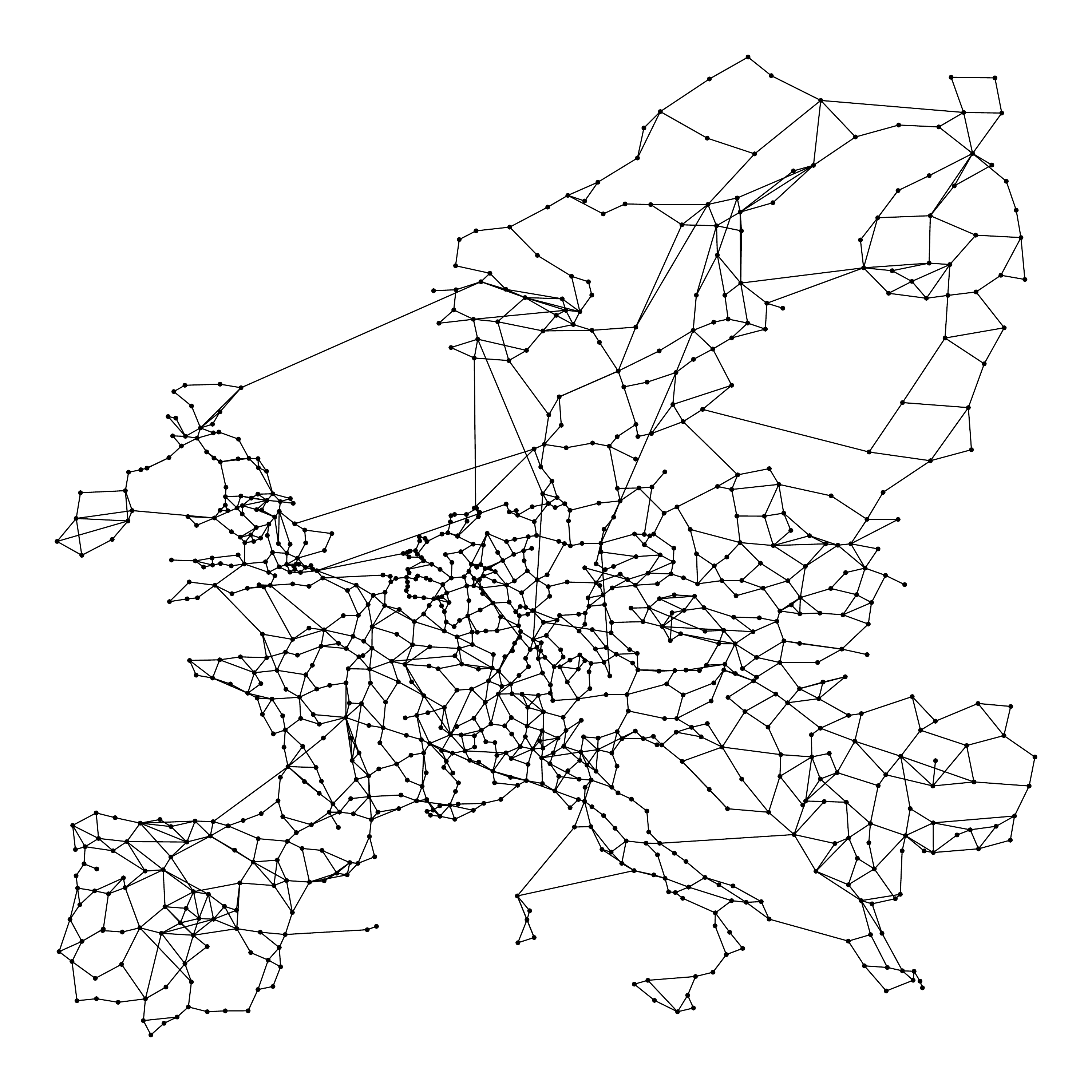}
	\caption{Sketch of an ENTSO-E network model with 1024 vertices and 1744 edges.}
	\label{fig:fig1}
\end{figure}
PyPSA-Eur~\cite{Hoersch2018} is an open model dataset of the European power system at the transmission network level that covers the full European Network of Transmission System Operators for Electricity (ENTSO-E) area (Fig.~\ref{fig:fig1}). 
We here considered pre-built networks~\cite{PyPSA-eur-data} (excluding the isolated power grids of Cyprus and Iceland)
that consist of $N \in \left\{37, 128, 256, 512, 1024\right\}$ vertices with $E \in \left\{77, 260, 483, 929, 1744\right\}$ edges, respectively.
Vertices represent buses to which consumers, generators or storages are connected directly or via lower-voltage distribution grids. 
Edges correspond to transmission lines or transformers that connect pairs of buses.\\

The classical IEEE test cases~\cite{Christie1999} consist of $N$ buses ($N \in \left\{14, 30, 57, 118, 300\right\}$) and represent a portion of the American Electric Power System as of the early 1960s (the 300-bus test case was developed in the early 1990s). 
The networks are partly synthetic and partly derived from real power grids and have $E \in \left\{20, 41, 78, 179, 409\right\}$ edges.

%%%%%%%%%%%%%%%%%%%%%%%%%%%%%%%%%%%%%%%%%%%%%%%%%%%%%%%%%%%%%%%%%%%%%%%%%%
\subsection{Network dynamics}
%%%%%%%%%%%%%%%%%%%%%%%%%%%%%%%%%%%%%%%%%%%%%%%%%%%%%%%%%%%%%%%%%%%%%%%%%%
%
We consider a generalized Kuramoto model~\cite{kuramoto1975,Guo2021}, which consists of an ensemble of $N$ coupled phase oscillators. 
Its evolution is governed by 
\begin{equation}
\dot{\theta_i}(t)=\omega_i+\kappa\sum_{j=1}^N A_{i,j} \sin{(\theta_j(t)-\theta_i(t))},
\label{eq:Kuramoto}
\end{equation}
where $\theta_i(t)$ is the instantaneous phase of the $i$th oscillator, and $\kappa$ denotes the global coupling strength.
$\mathbf{A} \in \left\{ 0,1 \right\}^{N \times N}$ is the symmetric adjacency matrix ($A_{ij} = A_{ji} = 1$, if and only if oscillators $i$ and $j$ are coupled) that represents a network.
We draw the oscillators' natural frequencies $\omega_i$ from a normal distribution $\mathcal{N}(2,0.1)$ with a mean of 2 and standard deviation $0.1$
to mimic fluctuations of the power grid frequency in a given network~\cite{Anvari2020,RydinGorjao2020,Schaefer2023}.

In our simulations, we vary $\kappa$ and generate $\numrea=100$ realizations of the network dynamics for each configuration of this control parameter.
$  $With equally distributed initial phases $\theta_i(0) \in [0,2\pi)$, redrawn natural frequencies, and after discarding $10^4$ transients, we generate synthetic phase time series of length $T = 10^4$ by numerically integrating~\cite{ansmann2018} Eq.~\ref{eq:Kuramoto} with a sampling interval of 1.

To characterize the collective dynamical behavior of a network, we employ the complex-valued order parameter
\begin{equation}
\widetilde{r}(t)=\frac{1}{N}\left|\sum_{j=1}^N e^{\iu\theta_j(t)}\right|,
\label{eq:orderparam}
\end{equation}
which takes on values between $0$ and $1$, where $1$ indicates complete phase synchronization.
Given the different number of vertices $N$ of the investigated networks, we derive an estimator~\cite{Kutil2012} for the order parameter as
\begin{align}
	r(t) = \sqrt{\frac{N}{N-1}\left( \widetilde{r}^2(t)-\frac{1}{N}\right)},
\end{align}
which allows an unbiased comparison between networks of different sizes.

%%%%%%%%%%%%%%%%%%%%%%%%%%%%%%%%%%%%%%%%%%%%%%%%%%%%%%%%%%%%%%%%%%%%%%%%%%
\subsection{Network metrics}
%%%%%%%%%%%%%%%%%%%%%%%%%%%%%%%%%%%%%%%%%%%%%%%%%%%%%%%%%%%%%%%%%%%%%%%%%%
%
We estimate relevant topological and spectral characteristics of the power grid models and compare them with those of paradigmatic network models~\cite{pagani2013,Rohden2014,Amani2021,Odor2022}, namely small-world networks~\cite{watts1998} (here with rewiring probability $p = 0.2$), scale-free networks~\cite{albert2002}, random networks\cite{erdos1959,Johnsonbaugh1991}, and 2D-lattices, whose respective numbers of vertices and edges compare to those of the power grid models.	

The global clustering coefficient $C$ assesses the tendency of vertices to cluster together, thus measuring the transitivity of a network.
It is defined as~\cite{Boccaletti2006}
\begin{align}
	C = \frac{3N_\Delta}{N_3},
\end{align}
where $N_\Delta$ is the number of closed triplets and $N_3$ the number of all triplets.\\

The average shortest path length $L$ assesses the average number of edges that must be traversed to reach any other vertex. 
$L$ is often associated with the efficiency of a network and is defined as~\cite{Boccaletti2006}
\begin{align}
	L = \frac{1}{N(N-1)}\sum_{i\neq j}d(i,j),
\end{align}
where ${(i,j)}\in\mathcal{V}$ and $d(i,j)$ is the length of the shortest path between vertices $i$ and $j$. 
$\mathcal{V}$ is the set of all vertices of a network.\\ 

Assortativity $A$ quantifies whether vertices preferentially connect to vertices with similar characteristics.
We here concentrate on the degree of a vertex, i.e., the number of edges it is incident to, and define $A$ as~\cite{Newman2003}
\begin{align}
	A = \frac{\sum_i e_{ii}-\sum_i a_i b_i}{1-\sum_i a_i b_i}.
\end{align}
Here $a_i = \sum_j e_{ij}$ and $b_j = \sum_i e_{ij}$, where $e_{ij}$ is the fraction of edges from a vertex of degree $i$ to a vertex of degree $j$.
Positive (negative) values of $A$ indicate an assortative (disassortative) network.
Disassortative networks are more vulnerable to perturbations and appear to be easier to synchronize than assortative networks~\cite{motter2005b,bernado2007}.\\

Synchronizability $S$ assesses the stability of the globally synchronized state of a network of coupled oscillators and is defined as the ratio of largest and smallest non-vanishing eigenvalue of the network Laplacian~\cite{pecora1998,barahona2002,donetti2005,atay2006}.
Given some vertex dynamics, the higher $S$ the less stable is the globally synchronized state of the network.

%%%%%%%%%%%%%%%%%%%%%%%%%%%%%%%%%%%%%%%%%%%%%%%%%%%%%%%%%%%%%%%%%%%%%%%%%%
\section{results}
\begin{figure}[htbp]
	\centering
	\includegraphics[width=0.5\textwidth]{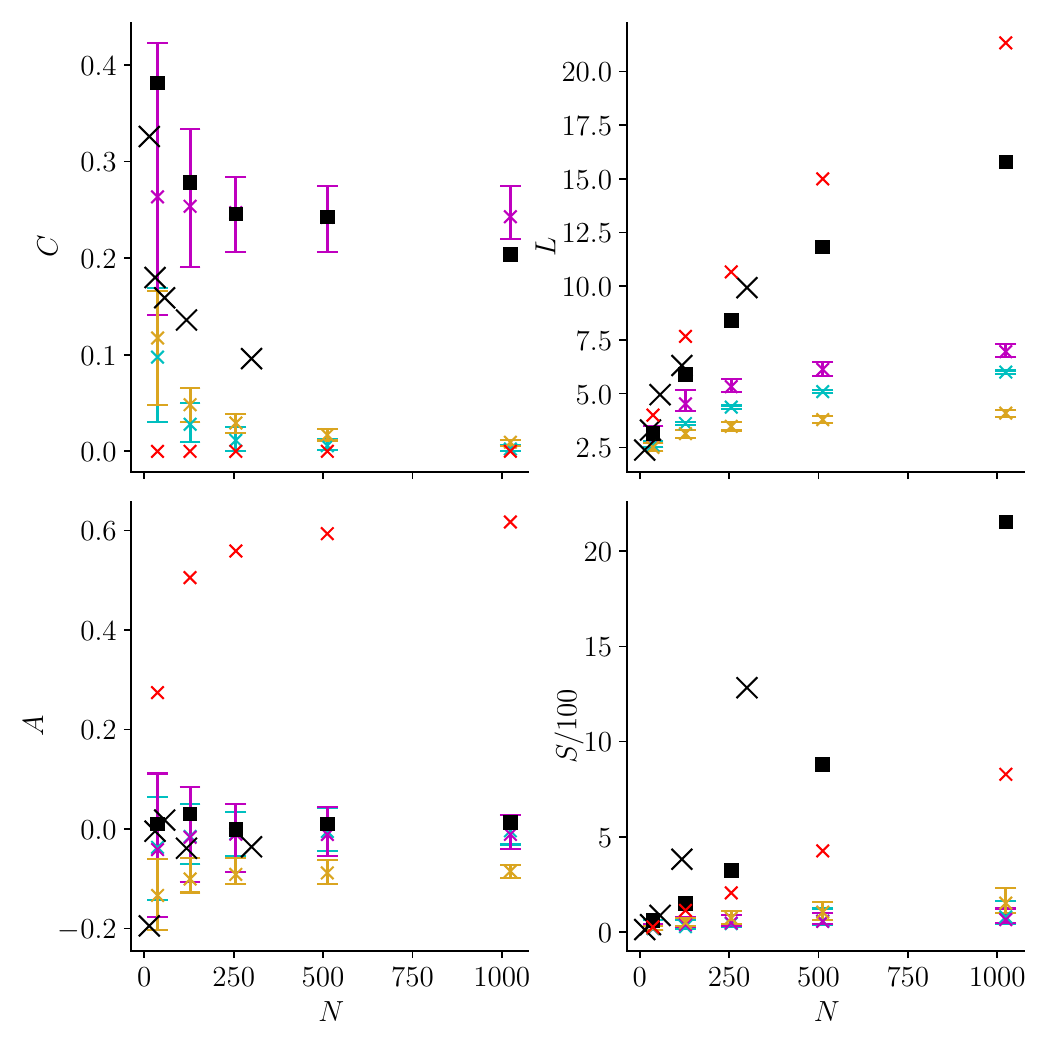}
	\caption{Topological and spectral characteristics of the ENTSO-E network models (black {\tiny $\blacksquare$}), the IEEE test cases (black $\times$), small-world (violet), scale-free (yellow), random (turquoise), and regular networks (2D-lattice; red) with different numbers of vertices $N$. 
		Global clustering coefficient $C$, average shortest path length $L$, assortativity $A$, and synchronizability $S$. 
		For the paradigmatic complex networks, we show the range (error bars) from 100 realizations of the networks.}
	\label{fig:fig2}
\end{figure}

%%%%%%%%%%%%%%%%%%%%%%%%%%%%%%%%%%%%%%%%%%%%%%%%%%%%%%%%%%%%%%%%%%%%%%%%%%
%
In Fig.~\ref{fig:fig2}, we compare the topological and spectral characteristics of the power grid models to those of the paradigmatic network models.

The global clustering coefficient $C$ of the ENTSO-E networks decreases with an increasing number of vertices, and this dependency compares to the one seen for the small-world networks. 
We observe a similar decrease for $C$ of the IEEE test cases. 
It attains, however, values much smaller than that of the ENTSO-E networks and appears to follow the dependency seen for the scale-free networks.
The average shortest path lengths $L$ of the power grids are quite similar and grow much faster with the number of vertices than the ones of the paradigmatic network models.
Assortativity $A$ of the power grids attains values close to 0 independent on the number of vertices $N$ and compares to the dependency seen for a random network.
Together, these findings indicate that the topologies of the investigated power grids cannot be clearly assigned to prototypical topologies~\citep{CotillaSanchez2012,pagani2013,Monfared2014,Espejo2018,Amani2021,Hartmann2021}.
They share, however, some commonalities in certain characteristics.

Interestingly, synchronizability $S$ of the power grids increases more strongly with network size $N$ than that of the paradigmatic network models. 
The comparably high values of $S$ of the power grids possibly indicate an unstable global synchronization state of oscillators coupled onto such networks.

\begin{figure*}[ht]
	\begin{center}
	\includegraphics[width=\textwidth]{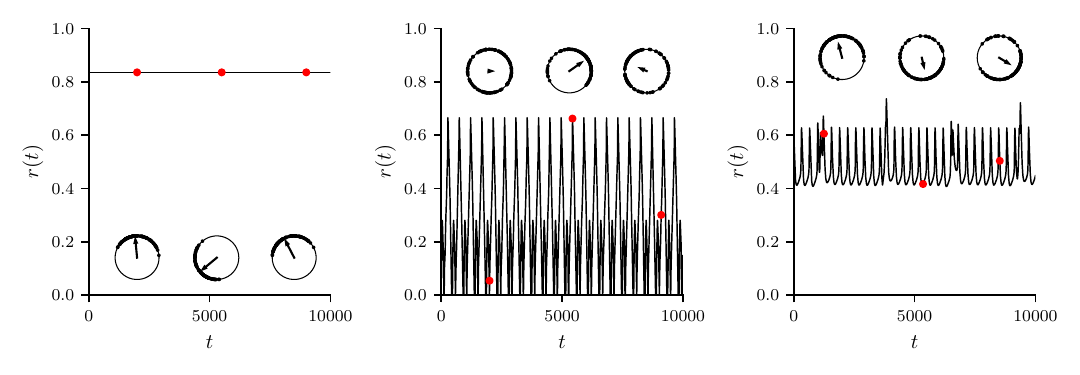}
	\caption{Examples of constant, periodic~\cite{ansmann2015b}, and non-periodic temporal evolutions of the order parameter. 
		ENTSO-E network model with $N = 128$ oscillators and coupling strength $\kappa = 0.3$.
		Insets show distribution of phases on the unit circle at times indicated by red dots. 
		The arrow points towards the mean phase and its length corresponds to the value of the order parameter.
		}
	\label{fig:fig3}
		\end{center}
\end{figure*}

%%%%%%%%%%%%%%%%%%%%%%%%%%%%%%%%%%%%
Considering the postulated impact of synchronizability on the synchronization dynamics of such oscillator networks, we investigate the oscillators' collective behavior on the power grid models and show in Fig.~\ref{fig:fig3} exemplary temporal evolutions of the order parameter for an ENTSO-E network model with $N=128$ oscillators.
We observe a variety of temporal evolutions depending on initial conditions, i.e., the oscillators' initial phases, as well as on the systems' disorder (drawn natural frequencies), but for a given coupling strength.
These range from constant to strictly periodic and to non-periodic, possibly chaotic~\cite{Bick2018} evolutions (see also Fig.~\ref{fig:fig8} in Appendix \ref{AppA}), and the majority of them indicate either no or partial phase-locking and only rarely full phase-locking.
Note that this variety holds independent on whether we fix the initial conditions or the systems' disorder.
We also note that such evolutions do not change upon increasing the phase time series length a hundred-fold.

\begin{figure}[htbp]
	\centering
	\includegraphics[width=0.45\textwidth]{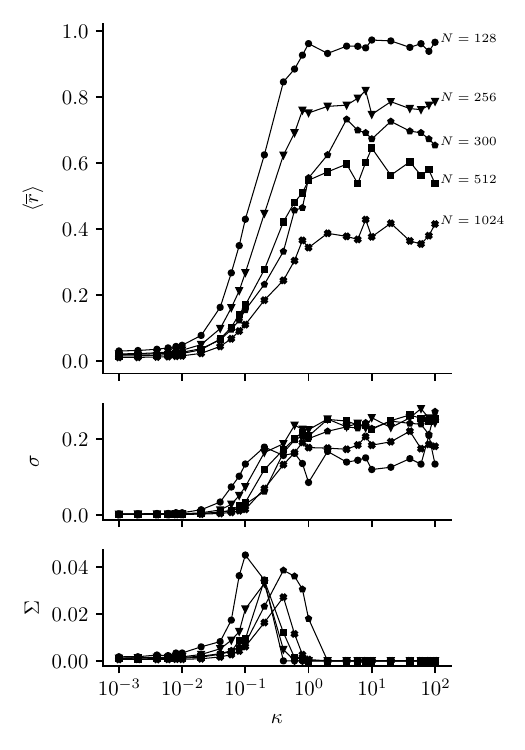}
	\caption{Impact of the coupling strength $\kappa$ on the order parameter (means ($\langle\overline{r}\rangle$) and standard deviations ($\sigma$ and $\Sigma$)) for ENTSO-E network models and the largest ($N=300$) of the IEEE test cases.
	Lines are for eye-guidance only.}
	\label{fig:fig4}
\end{figure}

Going beyond exemplary observations, we proceed with a more detailed characterization of the networks' synchronization dynamics.
To do so, we vary the coupling strength over five orders of magnitude ($\kappa \in \left[10^{-3},10^{2}\right]$) and estimate, for each value of $\kappa$, the long-time mean value of the order parameter, which we define as
\begin{align}
\nonumber\overline{r_i} &= \frac{1}{T}\sum_{t = 0}^{T}r_i(t)\\ 
\langle\overline{r}\rangle &= \frac{1}{\numrea}\sum_{i = 1}^{\numrea}\overline{r_i}.
\label{eq:meanr}
\end{align}
Moreover, we define the standard deviation of the mean values of the temporal means of each realization for the order parameter as
\begin{equation}
	\sigma = \textrm{std. dev.}\left( \overline{r_i} \right),
\end{equation}
and the standard deviation of the standard deviations of each realization $r_i(t)$ as
\begin{equation}
	\Sigma = \underset{i=1,\ldots,\numrea}{\textrm{std. dev.}} (\underset{t=1,\ldots,T}{\textrm{std. dev.}}(r_i(t))).
\end{equation}
Upon increasing the coupling strength $\kappa$, the mean order parameter (Eq.~\ref{eq:meanr}) increases sigmoidal-like, as expected~\cite{rodrigues2016} (Fig.~\ref{fig:fig4}).
However, only some of the realizations of the dynamics on the smallest investigated network ($N=128$) approach almost full phase-locking for sufficiently large coupling strength.
We observe a similar dependence on $\kappa$ for the standard deviation $\sigma$ of the mean values of the temporal means of the order parameter, which indicates that some coupling ($0.1\leq\kappa\leq1$) is required for non-trivial, non-constant evolutions of $r(t)$ to emerge. 
The dependence on the initial conditions and on systems' disorder (their combined impact is assessed with $\Sigma$), however, is most pronounced in an intermediate range of coupling strengths.
The observed increase in $\Sigma$ for moderate coupling can partly be attributed to the systems susceptibility to perturbations at the bifurcation point \cite{pikovsky2001,acebron2005}.

\begin{figure}[htbp]
	\centering
	\includegraphics[width=0.5\textwidth]{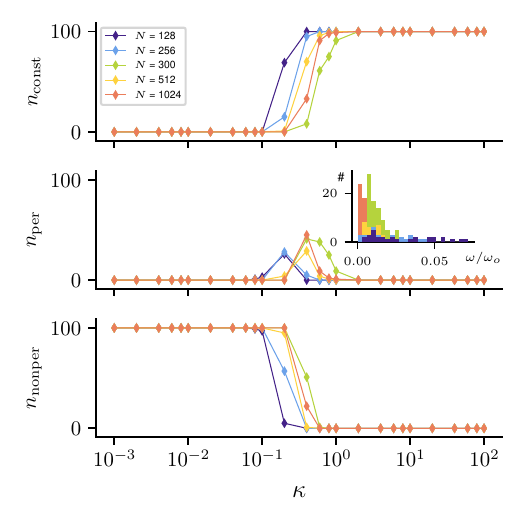}
	\caption{Number of occurrences of constant, periodic, and non-periodic temporal evolutions of the order parameter for $\numrea$ = 100 realizations of the network dynamics with different initial conditions of the oscillators but otherwise constant control parameters.
	Colors indicate numbers of vertices of the ENTSO-E network models and of the largest of the IEEE test cases ($N=300$).
	Lines are for eye-guidance only.
 	The inset shows the number of occurrences of frequencies $\omega$ of periodic evolutions normalized with respect to the mean $\omega_{\rm o}$ of the oscillators' natural frequencies. Frequencies of periodic evolutions are typically slow and are not natural multiples of the oscillators' natural frequencies.}
	\label{fig:fig5}
\end{figure}

Given these observations, we estimate the relative frequencies of constant, periodic, and non-periodic temporal evolutions of the order parameter for the aforementioned range of coupling strengths (Fig.~\ref{fig:fig5}).
For small couplings ($\kappa \lesssim 0.1$), we observe non-periodic temporal evolutions for all investigated networks, and for large couplings ($\kappa \gtrsim 1$), we solely find constant evolutions.
For an intermediate and quite narrow range of couplings ($0.1 \lesssim \kappa \lesssim 1$), however, we observe a quite sharp transition between preferentially non-periodic to preferentially constant temporal evolutions which is accompanied by the emergence of strictly periodic evolutions.
There is a tendency for this transition to occur at slightly smaller couplings for smaller networks sizes.
These observations allow for hypothesizing that other network properties may also impact on the synchronization dynamics. 
In the following, we therefore compare the synchronization dynamics on a selected power grid model (ENTSO-E; $N=256$) with those on paradigmatic network models.

\begin{figure}[htbp]
	\centering
	\includegraphics[width=0.45\textwidth]{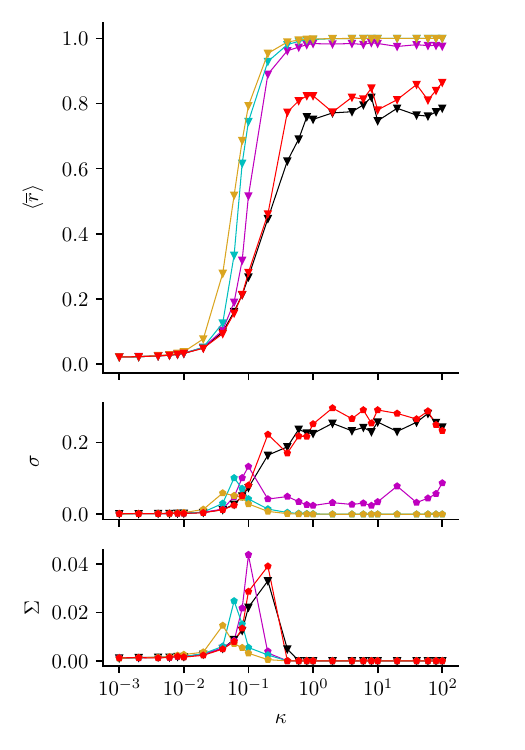}
	\caption{Same as Figure~\ref{fig:fig3} but for paradigmatic network topologies and a selected power grid model.
	Colors indicate network topologies (magenta: small-world, cyan: random, yellow: scale-free, red: 2D-lattice), and networks have $N=256$ vertices.
	Black color indicates the ENTSO-E network model of corresponding size (cf. Fig.~\ref{fig:fig4}).
	Lines are for eye-guidance only.}
	\label{fig:fig6}
\end{figure}

Just as for the mean order parameter $\langle\overline{r}\rangle$ in the case of the power grid models, $\langle\overline{r}\rangle$ in the case of the paradigmatic network topologies also increases sigmoidal-like with an increasing coupling strength $\kappa$ (Fig.~\ref{fig:fig6}).
However, for the complex topologies (small-world, random, and scale-free), $\langle\overline{r}\rangle$ indicates full phase-locking for large $\kappa$, while for the regular topology (2D-lattice) it does not.
Instead, $\langle\overline{r}\rangle$ levels off at a lower value of $\kappa$, similar to the case of the ENTSO-E network model of corresponding size.
For the complex topologies, the standard deviation $\sigma$ takes on largest values for an intermediate range of coupling strengths and is otherwise close to zero.
For the regular topology (2D-lattice), the increase of $\sigma$ with increasing $\kappa$ is again sigmoidal-like and comparable to the ENTSO-E case.
The impact of the initial conditions and the systems' disorder on the synchronization dynamics of the paradigmatic networks mimics the case of the power grid models and is also strongest in an intermediate range of couplings.

\begin{figure}[htbp]
	\centering
	\includegraphics[width=0.5\textwidth]{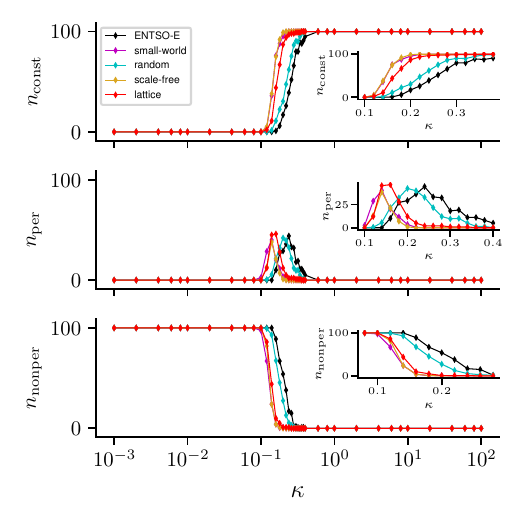}
	\caption{Number of occurrences of constant, periodic, and non-periodic temporal evolutions of the order parameter for $\numrea$ = 100 realizations of the paradigmatic networks with different initial conditions of the oscillators but otherwise constant control parameters.
	Colors indicate network topologies, and networks have $N=256$ vertices.
	Black color indicates the ENTSO-E network model of corresponding size (cf. Fig.~\ref{fig:fig5}).
	Lines are for eye-guidance only.}
	\label{fig:fig7}
\end{figure}

Figure~\ref{fig:fig7} summarizes our findings concerning the relative frequencies of constant, periodic, and non-periodic temporal evolutions of the order parameter depending on coupling strength.
Similar to the ENTSO-E case, we observe non-periodic temporal evolutions for all investigated networks for small coupling strengths ($\kappa \lesssim 0.1$) and exclusively constant evolutions for large coupling strengths($\kappa \gtrsim 0.4$).
The transition between the two types of evolutions is again confined to a narrow range of coupling strengths ($0.1 \lesssim \kappa \lesssim 0.4$) and is also accompanied by the emergence of strictly periodic evolutions.
Note that the transition occurs at similar coupling strengths for all paradigmatic topologies and the ENTSO-E network model of corresponding size, and for all networks we observe similar synchronization dynamics.
Interestingly, the impact of the coupling strength on the frequency of occurrence of constant, periodic, and non-periodic synchronization dynamics is comparable for random networks and the ENTSO-E network model. The other paradigmatic topologies including the 2D-lattice share a similar impact although for a different range of coupling strengths and at, on average, lower values of the coupling strength.

Due to their high synchronizability, the 2D-lattices and the ENTSO-E network models exhibit comparable synchronization dynamics, as indicated by their averaged order parameters $\langle\overline{r}\rangle$ (cf. Fig.~\ref{fig:fig6}).
These results align with our initial considerations on the impact of this spectral characteristic on synchronization dynamics.Yet, when considering the temporal evolutions of the order parameter, we find much more similar behavior between the seemingly opposing topologies of the random networks and the ENTSO-E network models (cf Fig.~\ref{fig:fig7}).

%%%%%%%%%%%%%%%%%%%%%%%%%%%%%%%%%%%%%%%%%%%%%%%%%%%%%%%%%%%%%%%%%%%%%%%%%%
\section{Conclusions}
%%%%%%%%%%%%%%%%%%%%%%%%%%%%%%%%%%%%%%%%%%%%%%%%%%%%%%%%%%%%%%%%%%%%%%%%%%

We investigated topological and spectral properties of models of European and US-American power grids in comparison to paradigmatic network models as well as their characteristics' implications for the synchronization dynamics of phase oscillators with heterogeneous natural frequencies coupled onto these networks.
Depending on the oscillators' initial conditions and on the systems' disorder, but for otherwise preset control parameters, the complex-valued order parameter exhibited temporal evolutions that ranged from constant to periodic and non-periodic, possibly chaotic.
The cycle durations of periodic evolutions point to emergent, possibly network-generated rhythms that are much slower than the natural periods of single oscillators.
Whether similar conclusions can also be drawn for non-periodic temporal evolutions remains to be investigated.

The synchronization dynamics on the power grid models compared largely to the ones seen for regular and random networks.
Relevant topological and spectral properties of the former, however, could not be clearly assigned to the properties of paradigmatic networks including other complex topologies (small-world and scale-free). 
Interestingly though, both topological and spectral properties of the power grid models point to a diminished capability of these networks to support a stable synchronization dynamics.
Likewise, this points to the possibility of improving the latter by modifying the networks' topology~\citep{wang2016c, tchuisseu2018}.

Instead of employing the long-time average of the complex-valued order parameter as a single measure for phase ordering, a more complete description of the systems' synchronization dynamics could possibly be achieved by investigating higher-order moments of the phase distribution~\citep{mardia2000,gong2019}.
These moments could also yield insights into the formation of stable and unstable clusters and their behavior over time as well as into their relationship to topological and spectral characteristics of the investigated networks.

%%%%%%%%%%%%%%%%%%%%%%%%%%%%%%%%%%%%%%%%%%%%%%%%%%%%%%%%%%%%%%%%%%%%%%%%%%
\begin{acknowledgments}
The authors acknowledge fruitful discussions with Mehrnaz Anvari and Leonardo Rydin Gorj\~{a}o.
\end{acknowledgments}

%%%%%%%%%%%%%%%%%%%%%%%%%%%%%%%%%%%%%%%%%%%%%%%%%%%%%%%%%%%%%%%%%%%%%%%%%%

\section*{AUTHOR DECLARATIONS}

\subsection*{Conflict of Interest}
The authors have no conflicts to disclose.

\subsection*{Data availability}
The data that support the findings of this study are available from the corresponding author upon reasonable request.

%%%%%%%%%%%%%%%%%%%%%%%%%%%%%%%%%%%%%%%%%%%%%%%%%%%%%%%%%%%%%%%%%%%%%%%%%%
\appendix
%\counterwithin{figure}{section}

%%%%%%%%%%%%%%%%%%%%%%%%%%%%%%%%%%%%%%%%%%%%%%%%%%%%%%%%%%%%%%%%%%%%%%%%%%
\section{Additional exemplary non-periodic synchronization dynamics on power grid models}
\label{AppA}
%%%%%%%%%%%%%%%%%%%%%%%%%%%%%%%%%%%%%%%%%%%%%%%%%%%%%%%%%%%%%%%%%%%%%%%%%%
%
\begin{figure}[htbp]
	\includegraphics[width=0.5\textwidth]{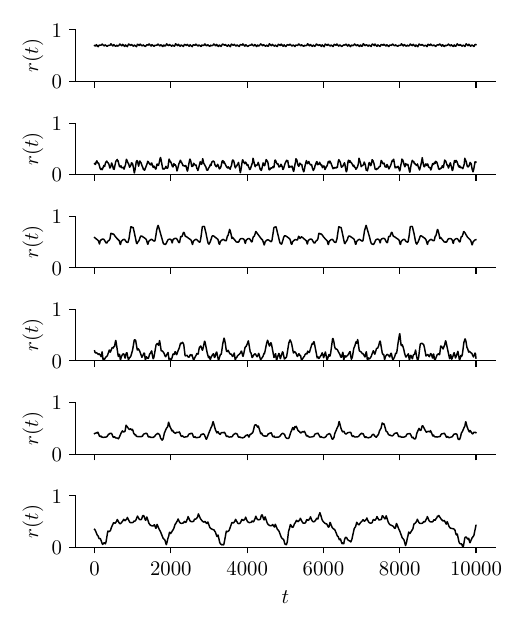}
	\caption{Examples of non-periodic temporal evolutions of the order parameter. 
		ENTSO-E network models with $N = 256$ oscillators, coupling strength $\kappa = 0.2$, and different initial conditions and systems' disorder (redrawn natural frequencies).}
\label{fig:fig8}
\end{figure}

%%%%%%%%%%%%%%%%%%%%%%%%%%%%%%%%%%%%%%%%%%%%%%%%%%%%%%%%%%%%%%%%%%%%%%%%%%

%\bibliography{newbib,bonnbib}

\begin{thebibliography}{85}%
\makeatletter
\providecommand \@ifxundefined [1]{%
 \@ifx{#1\undefined}
}%
\providecommand \@ifnum [1]{%
 \ifnum #1\expandafter \@firstoftwo
 \else \expandafter \@secondoftwo
 \fi
}%
\providecommand \@ifx [1]{%
 \ifx #1\expandafter \@firstoftwo
 \else \expandafter \@secondoftwo
 \fi
}%
\providecommand \natexlab [1]{#1}%
\providecommand \enquote  [1]{``#1''}%
\providecommand \bibnamefont  [1]{#1}%
\providecommand \bibfnamefont [1]{#1}%
\providecommand \citenamefont [1]{#1}%
\providecommand \href@noop [0]{\@secondoftwo}%
\providecommand \href [0]{\begingroup \@sanitize@url \@href}%
\providecommand \@href[1]{\@@startlink{#1}\@@href}%
\providecommand \@@href[1]{\endgroup#1\@@endlink}%
\providecommand \@sanitize@url [0]{\catcode `\\12\catcode `\$12\catcode
  `\&12\catcode `\#12\catcode `\^12\catcode `\_12\catcode `\%12\relax}%
\providecommand \@@startlink[1]{}%
\providecommand \@@endlink[0]{}%
\providecommand \url  [0]{\begingroup\@sanitize@url \@url }%
\providecommand \@url [1]{\endgroup\@href {#1}{\urlprefix }}%
\providecommand \urlprefix  [0]{URL }%
\providecommand \Eprint [0]{\href }%
\providecommand \doibase [0]{http://dx.doi.org/}%
\providecommand \selectlanguage [0]{\@gobble}%
\providecommand \bibinfo  [0]{\@secondoftwo}%
\providecommand \bibfield  [0]{\@secondoftwo}%
\providecommand \translation [1]{[#1]}%
\providecommand \BibitemOpen [0]{}%
\providecommand \bibitemStop [0]{}%
\providecommand \bibitemNoStop [0]{.\EOS\space}%
\providecommand \EOS [0]{\spacefactor3000\relax}%
\providecommand \BibitemShut  [1]{\csname bibitem#1\endcsname}%
\let\auto@bib@innerbib\@empty
%</preamble>
\bibitem [{\citenamefont {Pikovsky}, \citenamefont {Rosenblum},\ and\
  \citenamefont {Kurths}(2001)}]{pikovsky2001}%
  \BibitemOpen
  \bibfield  {author} {\bibinfo {author} {\bibfnamefont {A.~S.}\ \bibnamefont
  {Pikovsky}}, \bibinfo {author} {\bibfnamefont {M.~G.}\ \bibnamefont
  {Rosenblum}}, \ and\ \bibinfo {author} {\bibfnamefont {J.}~\bibnamefont
  {Kurths}},\ }\href {\doibase 10.1017/CBO9780511755743} {\emph {\bibinfo
  {title} {Synchronization: {A} universal concept in nonlinear sciences}}}\
  (\bibinfo  {publisher} {Cambridge University Press},\ \bibinfo {address}
  {Cambridge, UK},\ \bibinfo {year} {2001})\BibitemShut {NoStop}%
\bibitem [{\citenamefont {Glass}(2001)}]{glass2001}%
  \BibitemOpen
  \bibfield  {author} {\bibinfo {author} {\bibfnamefont {L.}~\bibnamefont
  {Glass}},\ }\bibfield  {title} {\enquote {\bibinfo {title} {Synchronization
  and rhythmic processes in physiology},}\ }\href@noop {} {\bibfield  {journal}
  {\bibinfo  {journal} {Nature}\ }\textbf {\bibinfo {volume} {410}},\ \bibinfo
  {pages} {277--284} (\bibinfo {year} {2001})}\BibitemShut {NoStop}%
\bibitem [{\citenamefont {Boccaletti}\ \emph {et~al.}(2002)\citenamefont
  {Boccaletti}, \citenamefont {Kurths}, \citenamefont {Osipov}, \citenamefont
  {Valladares},\ and\ \citenamefont {Zhou}}]{boccaletti2002}%
  \BibitemOpen
  \bibfield  {author} {\bibinfo {author} {\bibfnamefont {S.}~\bibnamefont
  {Boccaletti}}, \bibinfo {author} {\bibfnamefont {J.}~\bibnamefont {Kurths}},
  \bibinfo {author} {\bibfnamefont {G.}~\bibnamefont {Osipov}}, \bibinfo
  {author} {\bibfnamefont {D.~L.}\ \bibnamefont {Valladares}}, \ and\ \bibinfo
  {author} {\bibfnamefont {C.~S.}\ \bibnamefont {Zhou}},\ }\bibfield  {title}
  {\enquote {\bibinfo {title} {The synchronization of chaotic systems},}\
  }\href {\doibase 10.1016/S0370-1573(02)00137-0} {\bibfield  {journal}
  {\bibinfo  {journal} {Phys. Rep.}\ }\textbf {\bibinfo {volume} {366}},\
  \bibinfo {pages} {1--101} (\bibinfo {year} {2002})}\BibitemShut {NoStop}%
\bibitem [{\citenamefont {Mosekilde}, \citenamefont {Maistrenko},\ and\
  \citenamefont {Postnov}(2002)}]{mosekilde2002}%
  \BibitemOpen
  \bibfield  {author} {\bibinfo {author} {\bibfnamefont {E.}~\bibnamefont
  {Mosekilde}}, \bibinfo {author} {\bibfnamefont {Y.}~\bibnamefont
  {Maistrenko}}, \ and\ \bibinfo {author} {\bibfnamefont {D.}~\bibnamefont
  {Postnov}},\ }\href@noop {} {\emph {\bibinfo {title} {Chaotic
  synchronization: applications to living systems}}}\ (\bibinfo  {publisher}
  {World Scientific},\ \bibinfo {address} {Singapore},\ \bibinfo {year}
  {2002})\BibitemShut {NoStop}%
\bibitem [{\citenamefont {Rosenblum}\ and\ \citenamefont
  {Pikovsky}(2003)}]{Rosenblum2003}%
  \BibitemOpen
  \bibfield  {author} {\bibinfo {author} {\bibfnamefont {M.}~\bibnamefont
  {Rosenblum}}\ and\ \bibinfo {author} {\bibfnamefont {A.}~\bibnamefont
  {Pikovsky}},\ }\bibfield  {title} {\enquote {\bibinfo {title}
  {Synchronization: from pendulum clocks to chaotic lasers and chemical
  oscillators},}\ }\href@noop {} {\bibfield  {journal} {\bibinfo  {journal}
  {Contemp. Phys.}\ }\textbf {\bibinfo {volume} {44}},\ \bibinfo {pages}
  {401--416} (\bibinfo {year} {2003})}\BibitemShut {NoStop}%
\bibitem [{\citenamefont {Bennett}\ and\ \citenamefont
  {Zukin}(2004)}]{Bennett2004}%
  \BibitemOpen
  \bibfield  {author} {\bibinfo {author} {\bibfnamefont {M.~V.}\ \bibnamefont
  {Bennett}}\ and\ \bibinfo {author} {\bibfnamefont {R.~S.}\ \bibnamefont
  {Zukin}},\ }\bibfield  {title} {\enquote {\bibinfo {title} {Electrical
  coupling and neuronal synchronization in the mammalian brain},}\ }\href@noop
  {} {\bibfield  {journal} {\bibinfo  {journal} {Neuron}\ }\textbf {\bibinfo
  {volume} {41}},\ \bibinfo {pages} {495--511} (\bibinfo {year}
  {2004})}\BibitemShut {NoStop}%
\bibitem [{\citenamefont {Fell}\ and\ \citenamefont
  {Axmacher}(2011)}]{fell-axmacher2011}%
  \BibitemOpen
  \bibfield  {author} {\bibinfo {author} {\bibfnamefont {J.}~\bibnamefont
  {Fell}}\ and\ \bibinfo {author} {\bibfnamefont {N.}~\bibnamefont
  {Axmacher}},\ }\bibfield  {title} {\enquote {\bibinfo {title} {The role of
  phase synchronization in memory processes},}\ }\href@noop {} {\bibfield
  {journal} {\bibinfo  {journal} {Nat. Rev. Neurosci.}\ }\textbf {\bibinfo
  {volume} {12}},\ \bibinfo {pages} {105--118} (\bibinfo {year}
  {2011})}\BibitemShut {NoStop}%
\bibitem [{\citenamefont {Pikovsky}\ and\ \citenamefont
  {Rosenblum}(2015)}]{Pikovsky2015}%
  \BibitemOpen
  \bibfield  {author} {\bibinfo {author} {\bibfnamefont {A.}~\bibnamefont
  {Pikovsky}}\ and\ \bibinfo {author} {\bibfnamefont {M.}~\bibnamefont
  {Rosenblum}},\ }\bibfield  {title} {\enquote {\bibinfo {title} {Dynamics of
  globally coupled oscillators: {P}rogress and perspectives},}\ }\href@noop {}
  {\bibfield  {journal} {\bibinfo  {journal} {Chaos: An Interdisciplinary
  Journal of Nonlinear Science}\ }\textbf {\bibinfo {volume} {25}},\ \bibinfo
  {pages} {097616} (\bibinfo {year} {2015})}\BibitemShut {NoStop}%
\bibitem [{\citenamefont {Boccaletti}\ \emph {et~al.}(2018)\citenamefont
  {Boccaletti}, \citenamefont {Pisarchik}, \citenamefont {Del~Genio},\ and\
  \citenamefont {Amann}}]{Boccaletti2018}%
  \BibitemOpen
  \bibfield  {author} {\bibinfo {author} {\bibfnamefont {S.}~\bibnamefont
  {Boccaletti}}, \bibinfo {author} {\bibfnamefont {A.~N.}\ \bibnamefont
  {Pisarchik}}, \bibinfo {author} {\bibfnamefont {C.~I.}\ \bibnamefont
  {Del~Genio}}, \ and\ \bibinfo {author} {\bibfnamefont {A.}~\bibnamefont
  {Amann}},\ }\href@noop {} {\emph {\bibinfo {title} {Synchronization: from
  coupled systems to complex networks}}}\ (\bibinfo  {publisher} {Cambridge
  University Press},\ \bibinfo {address} {Cambridge, UK},\ \bibinfo {year}
  {2018})\BibitemShut {NoStop}%
\bibitem [{\citenamefont {Blekhman}(1988)}]{blekhman1988}%
  \BibitemOpen
  \bibfield  {author} {\bibinfo {author} {\bibfnamefont {I.~I.}\ \bibnamefont
  {Blekhman}},\ }\href@noop {} {\emph {\bibinfo {title} {Synchronization in
  science and technology}}}\ (\bibinfo  {publisher} {ASME Press},\ \bibinfo
  {year} {1988})\BibitemShut {NoStop}%
\bibitem [{\citenamefont {Nijmeijer}\ and\ \citenamefont
  {Rodriguez-Angeles}(2003)}]{Nijmeijer2003}%
  \BibitemOpen
  \bibfield  {author} {\bibinfo {author} {\bibfnamefont {H.}~\bibnamefont
  {Nijmeijer}}\ and\ \bibinfo {author} {\bibfnamefont {A.}~\bibnamefont
  {Rodriguez-Angeles}},\ }\href@noop {} {\emph {\bibinfo {title}
  {Synchronization of mechanical systems}}}\ (\bibinfo  {publisher} {World
  Scientific, Singapore},\ \bibinfo {year} {2003})\BibitemShut {NoStop}%
\bibitem [{\citenamefont {Kapitaniak}\ \emph {et~al.}(2012)\citenamefont
  {Kapitaniak}, \citenamefont {Czolczynski}, \citenamefont {Perlikowski},
  \citenamefont {Stefanski},\ and\ \citenamefont
  {Kapitaniak}}]{kapitaniak2012}%
  \BibitemOpen
  \bibfield  {author} {\bibinfo {author} {\bibfnamefont {M.}~\bibnamefont
  {Kapitaniak}}, \bibinfo {author} {\bibfnamefont {K.}~\bibnamefont
  {Czolczynski}}, \bibinfo {author} {\bibfnamefont {P.}~\bibnamefont
  {Perlikowski}}, \bibinfo {author} {\bibfnamefont {A.}~\bibnamefont
  {Stefanski}}, \ and\ \bibinfo {author} {\bibfnamefont {T.}~\bibnamefont
  {Kapitaniak}},\ }\bibfield  {title} {\enquote {\bibinfo {title}
  {Synchronization of clocks},}\ }\href {\doibase
  10.1016/j.physrep.2012.03.002} {\bibfield  {journal} {\bibinfo  {journal}
  {Phys. Rep.}\ }\textbf {\bibinfo {volume} {517}},\ \bibinfo {pages} {1--69}
  (\bibinfo {year} {2012})}\BibitemShut {NoStop}%
\bibitem [{\citenamefont {Motter}\ \emph {et~al.}(2013)\citenamefont {Motter},
  \citenamefont {Myers}, \citenamefont {Anghel},\ and\ \citenamefont
  {Nishikawa}}]{Motter2013}%
  \BibitemOpen
  \bibfield  {author} {\bibinfo {author} {\bibfnamefont {A.~E.}\ \bibnamefont
  {Motter}}, \bibinfo {author} {\bibfnamefont {S.~A.}\ \bibnamefont {Myers}},
  \bibinfo {author} {\bibfnamefont {M.}~\bibnamefont {Anghel}}, \ and\ \bibinfo
  {author} {\bibfnamefont {T.}~\bibnamefont {Nishikawa}},\ }\bibfield  {title}
  {\enquote {\bibinfo {title} {Spontaneous synchrony in power-grid networks},}\
  }\href@noop {} {\bibfield  {journal} {\bibinfo  {journal} {Nat. Phys.}\
  }\textbf {\bibinfo {volume} {9}},\ \bibinfo {pages} {191--197} (\bibinfo
  {year} {2013})}\BibitemShut {NoStop}%
\bibitem [{\citenamefont {D{\"o}rfler}\ and\ \citenamefont
  {Bullo}(2014)}]{Doerfler2014}%
  \BibitemOpen
  \bibfield  {author} {\bibinfo {author} {\bibfnamefont {F.}~\bibnamefont
  {D{\"o}rfler}}\ and\ \bibinfo {author} {\bibfnamefont {F.}~\bibnamefont
  {Bullo}},\ }\bibfield  {title} {\enquote {\bibinfo {title} {Synchronization
  in complex networks of phase oscillators: A survey},}\ }\href@noop {}
  {\bibfield  {journal} {\bibinfo  {journal} {Automatica}\ }\textbf {\bibinfo
  {volume} {50}},\ \bibinfo {pages} {1539--1564} (\bibinfo {year}
  {2014})}\BibitemShut {NoStop}%
\bibitem [{\citenamefont {Csaba}\ and\ \citenamefont
  {Porod}(2020)}]{Csaba2020}%
  \BibitemOpen
  \bibfield  {author} {\bibinfo {author} {\bibfnamefont {G.}~\bibnamefont
  {Csaba}}\ and\ \bibinfo {author} {\bibfnamefont {W.}~\bibnamefont {Porod}},\
  }\bibfield  {title} {\enquote {\bibinfo {title} {Coupled oscillators for
  computing: A review and perspective},}\ }\href@noop {} {\bibfield  {journal}
  {\bibinfo  {journal} {Appl. Phys. Rev.}\ }\textbf {\bibinfo {volume} {7}},\
  \bibinfo {pages} {011302} (\bibinfo {year} {2020})}\BibitemShut {NoStop}%
\bibitem [{\citenamefont {Witthaut}\ \emph {et~al.}(2022)\citenamefont
  {Witthaut}, \citenamefont {Hellmann}, \citenamefont {Kurths}, \citenamefont
  {Kettemann}, \citenamefont {Meyer-Ortmanns},\ and\ \citenamefont
  {Timme}}]{Witthaut2022}%
  \BibitemOpen
  \bibfield  {author} {\bibinfo {author} {\bibfnamefont {D.}~\bibnamefont
  {Witthaut}}, \bibinfo {author} {\bibfnamefont {F.}~\bibnamefont {Hellmann}},
  \bibinfo {author} {\bibfnamefont {J.}~\bibnamefont {Kurths}}, \bibinfo
  {author} {\bibfnamefont {S.}~\bibnamefont {Kettemann}}, \bibinfo {author}
  {\bibfnamefont {H.}~\bibnamefont {Meyer-Ortmanns}}, \ and\ \bibinfo {author}
  {\bibfnamefont {M.}~\bibnamefont {Timme}},\ }\bibfield  {title} {\enquote
  {\bibinfo {title} {Collective nonlinear dynamics and self-organization in
  decentralized power grids},}\ }\href@noop {} {\bibfield  {journal} {\bibinfo
  {journal} {Rev. Mod. Phys.}\ }\textbf {\bibinfo {volume} {94}},\ \bibinfo
  {pages} {015005} (\bibinfo {year} {2022})}\BibitemShut {NoStop}%
\bibitem [{\citenamefont {G\'omez-{Garde\~nes}}, \citenamefont {Moreno},\ and\
  \citenamefont {Arenas}(2007)}]{gomez2007}%
  \BibitemOpen
  \bibfield  {author} {\bibinfo {author} {\bibfnamefont {J.}~\bibnamefont
  {G\'omez-{Garde\~nes}}}, \bibinfo {author} {\bibfnamefont {Y.}~\bibnamefont
  {Moreno}}, \ and\ \bibinfo {author} {\bibfnamefont {A.}~\bibnamefont
  {Arenas}},\ }\bibfield  {title} {\enquote {\bibinfo {title} {Paths to
  synchronization on complex networks},}\ }\href {\doibase
  10.1103/PhysRevLett.98.034101} {\bibfield  {journal} {\bibinfo  {journal}
  {Phys. Rev. Lett.}\ }\textbf {\bibinfo {volume} {98}},\ \bibinfo {pages}
  {034101} (\bibinfo {year} {2007})}\BibitemShut {NoStop}%
\bibitem [{\citenamefont {Gomez-Gardenes}, \citenamefont {Moreno},\ and\
  \citenamefont {Arenas}(2007)}]{gomez2007b}%
  \BibitemOpen
  \bibfield  {author} {\bibinfo {author} {\bibfnamefont {J.}~\bibnamefont
  {Gomez-Gardenes}}, \bibinfo {author} {\bibfnamefont {Y.}~\bibnamefont
  {Moreno}}, \ and\ \bibinfo {author} {\bibfnamefont {A.}~\bibnamefont
  {Arenas}},\ }\bibfield  {title} {\enquote {\bibinfo {title}
  {Synchronizability determined by coupling strengths and topology on complex
  networks},}\ }\href {\doibase 10.1103/PhysRevE.75.066106} {\bibfield
  {journal} {\bibinfo  {journal} {Phys. Rev. E}\ }\textbf {\bibinfo {volume}
  {75}},\ \bibinfo {pages} {066106} (\bibinfo {year} {2007})}\BibitemShut
  {NoStop}%
\bibitem [{\citenamefont {Arenas}\ \emph {et~al.}(2008)\citenamefont {Arenas},
  \citenamefont {D\'iaz-Guilera}, \citenamefont {Kurths}, \citenamefont
  {Moreno},\ and\ \citenamefont {Zhou}}]{arenas2008}%
  \BibitemOpen
  \bibfield  {author} {\bibinfo {author} {\bibfnamefont {A.}~\bibnamefont
  {Arenas}}, \bibinfo {author} {\bibfnamefont {A.}~\bibnamefont
  {D\'iaz-Guilera}}, \bibinfo {author} {\bibfnamefont {J.}~\bibnamefont
  {Kurths}}, \bibinfo {author} {\bibfnamefont {Y.}~\bibnamefont {Moreno}}, \
  and\ \bibinfo {author} {\bibfnamefont {C.}~\bibnamefont {Zhou}},\ }\bibfield
  {title} {\enquote {\bibinfo {title} {Synchronization in complex networks},}\
  }\href {\doibase 10.1016/j.physrep.2008.09.002} {\bibfield  {journal}
  {\bibinfo  {journal} {Phys. Rep.}\ }\textbf {\bibinfo {volume} {469}},\
  \bibinfo {pages} {93--153} (\bibinfo {year} {2008})}\BibitemShut {NoStop}%
\bibitem [{\citenamefont {Rohden}\ \emph {et~al.}(2014)\citenamefont {Rohden},
  \citenamefont {Sorge}, \citenamefont {Witthaut},\ and\ \citenamefont
  {Timme}}]{Rohden2014}%
  \BibitemOpen
  \bibfield  {author} {\bibinfo {author} {\bibfnamefont {M.}~\bibnamefont
  {Rohden}}, \bibinfo {author} {\bibfnamefont {A.}~\bibnamefont {Sorge}},
  \bibinfo {author} {\bibfnamefont {D.}~\bibnamefont {Witthaut}}, \ and\
  \bibinfo {author} {\bibfnamefont {M.}~\bibnamefont {Timme}},\ }\bibfield
  {title} {\enquote {\bibinfo {title} {Impact of network topology on synchrony
  of oscillatory power grids},}\ }\href@noop {} {\bibfield  {journal} {\bibinfo
   {journal} {Chaos}\ }\textbf {\bibinfo {volume} {24}},\ \bibinfo {pages}
  {013123} (\bibinfo {year} {2014})}\BibitemShut {NoStop}%
\bibitem [{\citenamefont {Strogatz}(2000)}]{strogatz2000}%
  \BibitemOpen
  \bibfield  {author} {\bibinfo {author} {\bibfnamefont {S.~H.}\ \bibnamefont
  {Strogatz}},\ }\bibfield  {title} {\enquote {\bibinfo {title} {From
  {Kuramoto} to {Crawford}: exploring the onset of synchronization in
  populations of coupled oscillators},}\ }\href@noop {} {\bibfield  {journal}
  {\bibinfo  {journal} {Physica~D}\ }\textbf {\bibinfo {volume} {143}},\
  \bibinfo {pages} {1--20} (\bibinfo {year} {2000})}\BibitemShut {NoStop}%
\bibitem [{\citenamefont {Acebr\'on}\ \emph {et~al.}(2005)\citenamefont
  {Acebr\'on}, \citenamefont {Bonilla}, \citenamefont {{P\'erez Vicente}},
  \citenamefont {Ritort},\ and\ \citenamefont {Spigler}}]{acebron2005}%
  \BibitemOpen
  \bibfield  {author} {\bibinfo {author} {\bibfnamefont {J.~A.}\ \bibnamefont
  {Acebr\'on}}, \bibinfo {author} {\bibfnamefont {L.~L.}\ \bibnamefont
  {Bonilla}}, \bibinfo {author} {\bibfnamefont {C.~J.}\ \bibnamefont {{P\'erez
  Vicente}}}, \bibinfo {author} {\bibfnamefont {F.}~\bibnamefont {Ritort}}, \
  and\ \bibinfo {author} {\bibfnamefont {R.}~\bibnamefont {Spigler}},\
  }\bibfield  {title} {\enquote {\bibinfo {title} {The {K}uramoto model: A
  simple paradigm for synchronization phenomena},}\ }\href@noop {} {\bibfield
  {journal} {\bibinfo  {journal} {Rev. Mod. Phys.}\ }\textbf {\bibinfo {volume}
  {77}},\ \bibinfo {pages} {137--185} (\bibinfo {year} {2005})}\BibitemShut
  {NoStop}%
\bibitem [{\citenamefont {Breakspear}, \citenamefont {Heitmann},\ and\
  \citenamefont {Daffertshofer}(2010)}]{Breakspear2010}%
  \BibitemOpen
  \bibfield  {author} {\bibinfo {author} {\bibfnamefont {M.}~\bibnamefont
  {Breakspear}}, \bibinfo {author} {\bibfnamefont {S.}~\bibnamefont
  {Heitmann}}, \ and\ \bibinfo {author} {\bibfnamefont {A.}~\bibnamefont
  {Daffertshofer}},\ }\bibfield  {title} {\enquote {\bibinfo {title}
  {Generative models of cortical oscillations: neurobiological implications of
  the {Kuramoto} model},}\ }\href@noop {} {\bibfield  {journal} {\bibinfo
  {journal} {Front. Human Neurosci.}\ }\textbf {\bibinfo {volume} {4}},\
  \bibinfo {pages} {190} (\bibinfo {year} {2010})}\BibitemShut {NoStop}%
\bibitem [{\citenamefont {Rodrigues}\ \emph {et~al.}(2016)\citenamefont
  {Rodrigues}, \citenamefont {Peron}, \citenamefont {Ji},\ and\ \citenamefont
  {Kurths}}]{rodrigues2016}%
  \BibitemOpen
  \bibfield  {author} {\bibinfo {author} {\bibfnamefont {F.~A.}\ \bibnamefont
  {Rodrigues}}, \bibinfo {author} {\bibfnamefont {T.~K.~D.}\ \bibnamefont
  {Peron}}, \bibinfo {author} {\bibfnamefont {P.}~\bibnamefont {Ji}}, \ and\
  \bibinfo {author} {\bibfnamefont {J.}~\bibnamefont {Kurths}},\ }\bibfield
  {title} {\enquote {\bibinfo {title} {The {Kuramoto} model in complex
  networks},}\ }\href {\doibase 10.1016/j.physrep.2015.10.008} {\bibfield
  {journal} {\bibinfo  {journal} {Phys. Rep.}\ }\textbf {\bibinfo {volume}
  {610}},\ \bibinfo {pages} {1--98} (\bibinfo {year} {2016})}\BibitemShut
  {NoStop}%
\bibitem [{\citenamefont {Bick}\ \emph {et~al.}(2020)\citenamefont {Bick},
  \citenamefont {Goodfellow}, \citenamefont {Laing},\ and\ \citenamefont
  {Martens}}]{Bick2020}%
  \BibitemOpen
  \bibfield  {author} {\bibinfo {author} {\bibfnamefont {C.}~\bibnamefont
  {Bick}}, \bibinfo {author} {\bibfnamefont {M.}~\bibnamefont {Goodfellow}},
  \bibinfo {author} {\bibfnamefont {C.~R.}\ \bibnamefont {Laing}}, \ and\
  \bibinfo {author} {\bibfnamefont {E.~A.}\ \bibnamefont {Martens}},\
  }\bibfield  {title} {\enquote {\bibinfo {title} {Understanding the dynamics
  of biological and neural oscillator networks through exact mean-field
  reductions: a review},}\ }\href@noop {} {\bibfield  {journal} {\bibinfo
  {journal} {J. Math. Neurosci.}\ }\textbf {\bibinfo {volume} {10}},\ \bibinfo
  {pages} {9} (\bibinfo {year} {2020})}\BibitemShut {NoStop}%
\bibitem [{\citenamefont {Kuramoto}(1984)}]{kuramoto1984}%
  \BibitemOpen
  \bibfield  {author} {\bibinfo {author} {\bibfnamefont {Y.}~\bibnamefont
  {Kuramoto}},\ }\href@noop {} {\emph {\bibinfo {title} {Chemical Oscillations,
  Waves and Turbulence}}}\ (\bibinfo  {publisher} {Springer Verlag},\ \bibinfo
  {address} {Berlin},\ \bibinfo {year} {1984})\BibitemShut {NoStop}%
\bibitem [{\citenamefont {Schr{\"o}der}, \citenamefont {Timme},\ and\
  \citenamefont {Witthaut}(2017)}]{Schroeder2017}%
  \BibitemOpen
  \bibfield  {author} {\bibinfo {author} {\bibfnamefont {M.}~\bibnamefont
  {Schr{\"o}der}}, \bibinfo {author} {\bibfnamefont {M.}~\bibnamefont {Timme}},
  \ and\ \bibinfo {author} {\bibfnamefont {D.}~\bibnamefont {Witthaut}},\
  }\bibfield  {title} {\enquote {\bibinfo {title} {A universal order parameter
  for synchrony in networks of limit cycle oscillators},}\ }\href@noop {}
  {\bibfield  {journal} {\bibinfo  {journal} {Chaos: An Interdisciplinary
  Journal of Nonlinear Science}\ }\textbf {\bibinfo {volume} {27}} (\bibinfo
  {year} {2017})}\BibitemShut {NoStop}%
\bibitem [{\citenamefont {Ott}\ and\ \citenamefont {Antonsen}(2009)}]{Ott2009}%
  \BibitemOpen
  \bibfield  {author} {\bibinfo {author} {\bibfnamefont {E.}~\bibnamefont
  {Ott}}\ and\ \bibinfo {author} {\bibfnamefont {T.~M.}\ \bibnamefont
  {Antonsen}},\ }\bibfield  {title} {\enquote {\bibinfo {title} {Long time
  evolution of phase oscillator systems},}\ }\href@noop {} {\bibfield
  {journal} {\bibinfo  {journal} {Chaos: An interdisciplinary journal of
  nonlinear science}\ }\textbf {\bibinfo {volume} {19}},\ \bibinfo {pages}
  {023117} (\bibinfo {year} {2009})}\BibitemShut {NoStop}%
\bibitem [{\citenamefont {Mirollo}(2012)}]{Mirollo2012}%
  \BibitemOpen
  \bibfield  {author} {\bibinfo {author} {\bibfnamefont {R.~E.}\ \bibnamefont
  {Mirollo}},\ }\bibfield  {title} {\enquote {\bibinfo {title} {The asymptotic
  behavior of the order parameter for the infinite-{N} {K}uramoto model},}\
  }\href@noop {} {\bibfield  {journal} {\bibinfo  {journal} {Chaos: An
  Interdisciplinary Journal of Nonlinear Science}\ }\textbf {\bibinfo {volume}
  {22}},\ \bibinfo {pages} {033133} (\bibinfo {year} {2012})}\BibitemShut
  {NoStop}%
\bibitem [{\citenamefont {Bick}, \citenamefont {Panaggio},\ and\ \citenamefont
  {Martens}(2018)}]{Bick2018}%
  \BibitemOpen
  \bibfield  {author} {\bibinfo {author} {\bibfnamefont {C.}~\bibnamefont
  {Bick}}, \bibinfo {author} {\bibfnamefont {M.~J.}\ \bibnamefont {Panaggio}},
  \ and\ \bibinfo {author} {\bibfnamefont {E.~A.}\ \bibnamefont {Martens}},\
  }\bibfield  {title} {\enquote {\bibinfo {title} {Chaos in {Kuramoto}
  oscillator networks},}\ }\href@noop {} {\bibfield  {journal} {\bibinfo
  {journal} {Chaos: An Interdisciplinary Journal of Nonlinear Science}\
  }\textbf {\bibinfo {volume} {28}},\ \bibinfo {pages} {071102} (\bibinfo
  {year} {2018})}\BibitemShut {NoStop}%
\bibitem [{\citenamefont {Smith}\ and\ \citenamefont
  {Gottwald}(2019)}]{smith2019}%
  \BibitemOpen
  \bibfield  {author} {\bibinfo {author} {\bibfnamefont {L.~D.}\ \bibnamefont
  {Smith}}\ and\ \bibinfo {author} {\bibfnamefont {G.~A.}\ \bibnamefont
  {Gottwald}},\ }\bibfield  {title} {\enquote {\bibinfo {title} {Chaos in
  networks of coupled oscillators with multimodal natural frequency
  distributions},}\ }\href@noop {} {\bibfield  {journal} {\bibinfo  {journal}
  {Chaos: An Interdisciplinary Journal of Nonlinear Science}\ }\textbf
  {\bibinfo {volume} {29}},\ \bibinfo {pages} {093127} (\bibinfo {year}
  {2019})}\BibitemShut {NoStop}%
\bibitem [{\citenamefont {Clusella}\ and\ \citenamefont
  {Politi}(2020)}]{clusella2020}%
  \BibitemOpen
  \bibfield  {author} {\bibinfo {author} {\bibfnamefont {P.}~\bibnamefont
  {Clusella}}\ and\ \bibinfo {author} {\bibfnamefont {A.}~\bibnamefont
  {Politi}},\ }\bibfield  {title} {\enquote {\bibinfo {title} {Irregular
  collective dynamics in a {Kuramoto--Daido} system},}\ }\href@noop {}
  {\bibfield  {journal} {\bibinfo  {journal} {J. Phys. Complex.}\ }\textbf
  {\bibinfo {volume} {2}},\ \bibinfo {pages} {014002} (\bibinfo {year}
  {2020})}\BibitemShut {NoStop}%
\bibitem [{\citenamefont {Yeung}\ and\ \citenamefont
  {Strogatz}(1999)}]{Yeung1999}%
  \BibitemOpen
  \bibfield  {author} {\bibinfo {author} {\bibfnamefont {M.~S.}\ \bibnamefont
  {Yeung}}\ and\ \bibinfo {author} {\bibfnamefont {S.~H.}\ \bibnamefont
  {Strogatz}},\ }\bibfield  {title} {\enquote {\bibinfo {title} {Time delay in
  the {K}uramoto model of coupled oscillators},}\ }\href@noop {} {\bibfield
  {journal} {\bibinfo  {journal} {Phys. Rev. lett.}\ }\textbf {\bibinfo
  {volume} {82}},\ \bibinfo {pages} {648} (\bibinfo {year} {1999})}\BibitemShut
  {NoStop}%
\bibitem [{\citenamefont {Senthilkumar}\ \emph {et~al.}(2011)\citenamefont
  {Senthilkumar}, \citenamefont {Suresh}, \citenamefont {Sheeba}, \citenamefont
  {Lakshmanan},\ and\ \citenamefont {Kurths}}]{Senthilkumar2011}%
  \BibitemOpen
  \bibfield  {author} {\bibinfo {author} {\bibfnamefont {D.}~\bibnamefont
  {Senthilkumar}}, \bibinfo {author} {\bibfnamefont {R.}~\bibnamefont
  {Suresh}}, \bibinfo {author} {\bibfnamefont {J.~H.}\ \bibnamefont {Sheeba}},
  \bibinfo {author} {\bibfnamefont {M.}~\bibnamefont {Lakshmanan}}, \ and\
  \bibinfo {author} {\bibfnamefont {J.}~\bibnamefont {Kurths}},\ }\bibfield
  {title} {\enquote {\bibinfo {title} {Delay-enhanced coherent chaotic
  oscillations in networks with large disorders},}\ }\href@noop {} {\bibfield
  {journal} {\bibinfo  {journal} {Phys. Rev. E}\ }\textbf {\bibinfo {volume}
  {84}},\ \bibinfo {pages} {066206} (\bibinfo {year} {2011})}\BibitemShut
  {NoStop}%
\bibitem [{\citenamefont {Gjurchinovski}, \citenamefont {Sch{\"o}ll},\ and\
  \citenamefont {Zakharova}(2017)}]{Gjurchinovski2017}%
  \BibitemOpen
  \bibfield  {author} {\bibinfo {author} {\bibfnamefont {A.}~\bibnamefont
  {Gjurchinovski}}, \bibinfo {author} {\bibfnamefont {E.}~\bibnamefont
  {Sch{\"o}ll}}, \ and\ \bibinfo {author} {\bibfnamefont {A.}~\bibnamefont
  {Zakharova}},\ }\bibfield  {title} {\enquote {\bibinfo {title} {Control of
  amplitude chimeras by time delay in oscillator networks},}\ }\href@noop {}
  {\bibfield  {journal} {\bibinfo  {journal} {Phys. Rev. E}\ }\textbf {\bibinfo
  {volume} {95}},\ \bibinfo {pages} {042218} (\bibinfo {year}
  {2017})}\BibitemShut {NoStop}%
\bibitem [{\citenamefont {Bick}\ \emph {et~al.}(2011)\citenamefont {Bick},
  \citenamefont {Timme}, \citenamefont {Paulikat}, \citenamefont {Rathlev},\
  and\ \citenamefont {Ashwin}}]{Bick2011}%
  \BibitemOpen
  \bibfield  {author} {\bibinfo {author} {\bibfnamefont {C.}~\bibnamefont
  {Bick}}, \bibinfo {author} {\bibfnamefont {M.}~\bibnamefont {Timme}},
  \bibinfo {author} {\bibfnamefont {D.}~\bibnamefont {Paulikat}}, \bibinfo
  {author} {\bibfnamefont {D.}~\bibnamefont {Rathlev}}, \ and\ \bibinfo
  {author} {\bibfnamefont {P.}~\bibnamefont {Ashwin}},\ }\bibfield  {title}
  {\enquote {\bibinfo {title} {Chaos in symmetric phase oscillator networks},}\
  }\href@noop {} {\bibfield  {journal} {\bibinfo  {journal} {Phys. Rev. Lett.}\
  }\textbf {\bibinfo {volume} {107}},\ \bibinfo {pages} {244101} (\bibinfo
  {year} {2011})}\BibitemShut {NoStop}%
\bibitem [{\citenamefont {Labavi{\'c}}\ and\ \citenamefont
  {Meyer-Ortmanns}(2017)}]{Labavic2017}%
  \BibitemOpen
  \bibfield  {author} {\bibinfo {author} {\bibfnamefont {D.}~\bibnamefont
  {Labavi{\'c}}}\ and\ \bibinfo {author} {\bibfnamefont {H.}~\bibnamefont
  {Meyer-Ortmanns}},\ }\bibfield  {title} {\enquote {\bibinfo {title}
  {Long-period clocks from short-period oscillators},}\ }\href@noop {}
  {\bibfield  {journal} {\bibinfo  {journal} {Chaos: An Interdisciplinary
  Journal of Nonlinear Science}\ }\textbf {\bibinfo {volume} {27}},\ \bibinfo
  {pages} {083103} (\bibinfo {year} {2017})}\BibitemShut {NoStop}%
\bibitem [{\citenamefont {Chouzouris}\ \emph {et~al.}(2018)\citenamefont
  {Chouzouris}, \citenamefont {Omelchenko}, \citenamefont {Zakharova},
  \citenamefont {Hlinka}, \citenamefont {Jiruska},\ and\ \citenamefont
  {Sch{\"o}ll}}]{Chouzouris2018}%
  \BibitemOpen
  \bibfield  {author} {\bibinfo {author} {\bibfnamefont {T.}~\bibnamefont
  {Chouzouris}}, \bibinfo {author} {\bibfnamefont {I.}~\bibnamefont
  {Omelchenko}}, \bibinfo {author} {\bibfnamefont {A.}~\bibnamefont
  {Zakharova}}, \bibinfo {author} {\bibfnamefont {J.}~\bibnamefont {Hlinka}},
  \bibinfo {author} {\bibfnamefont {P.}~\bibnamefont {Jiruska}}, \ and\
  \bibinfo {author} {\bibfnamefont {E.}~\bibnamefont {Sch{\"o}ll}},\ }\bibfield
   {title} {\enquote {\bibinfo {title} {Chimera states in brain networks:
  Empirical neural vs. modular fractal connectivity},}\ }\href@noop {}
  {\bibfield  {journal} {\bibinfo  {journal} {Chaos: An Interdisciplinary
  Journal of Nonlinear Science}\ }\textbf {\bibinfo {volume} {28}},\ \bibinfo
  {pages} {045112} (\bibinfo {year} {2018})}\BibitemShut {NoStop}%
\bibitem [{\citenamefont {Paolini}\ \emph {et~al.}(2022)\citenamefont
  {Paolini}, \citenamefont {Ciszak}, \citenamefont {Marino}, \citenamefont
  {Olmi},\ and\ \citenamefont {Torcini}}]{Paolini2022}%
  \BibitemOpen
  \bibfield  {author} {\bibinfo {author} {\bibfnamefont {G.}~\bibnamefont
  {Paolini}}, \bibinfo {author} {\bibfnamefont {M.}~\bibnamefont {Ciszak}},
  \bibinfo {author} {\bibfnamefont {F.}~\bibnamefont {Marino}}, \bibinfo
  {author} {\bibfnamefont {S.}~\bibnamefont {Olmi}}, \ and\ \bibinfo {author}
  {\bibfnamefont {A.}~\bibnamefont {Torcini}},\ }\bibfield  {title} {\enquote
  {\bibinfo {title} {Collective excitability in highly diluted random networks
  of oscillators},}\ }\href@noop {} {\bibfield  {journal} {\bibinfo  {journal}
  {Chaos: An Interdisciplinary Journal of Nonlinear Science}\ }\textbf
  {\bibinfo {volume} {32}},\ \bibinfo {pages} {103108} (\bibinfo {year}
  {2022})}\BibitemShut {NoStop}%
\bibitem [{\citenamefont {Tadi{\'c}}, \citenamefont {Chutani},\ and\
  \citenamefont {Gupte}(2022)}]{Tadic2022}%
  \BibitemOpen
  \bibfield  {author} {\bibinfo {author} {\bibfnamefont {B.}~\bibnamefont
  {Tadi{\'c}}}, \bibinfo {author} {\bibfnamefont {M.}~\bibnamefont {Chutani}},
  \ and\ \bibinfo {author} {\bibfnamefont {N.}~\bibnamefont {Gupte}},\
  }\bibfield  {title} {\enquote {\bibinfo {title} {Multiscale fractality in
  partial phase synchronisation on simplicial complexes around brain hubs},}\
  }\href@noop {} {\bibfield  {journal} {\bibinfo  {journal} {Chaos, Solitons \&
  Fractals}\ }\textbf {\bibinfo {volume} {160}},\ \bibinfo {pages} {112201}
  (\bibinfo {year} {2022})}\BibitemShut {NoStop}%
\bibitem [{\citenamefont {Buscarino}\ \emph {et~al.}(2015)\citenamefont
  {Buscarino}, \citenamefont {Frasca}, \citenamefont {Gambuzza},\ and\
  \citenamefont {H{\"o}vel}}]{Buscarino2015}%
  \BibitemOpen
  \bibfield  {author} {\bibinfo {author} {\bibfnamefont {A.}~\bibnamefont
  {Buscarino}}, \bibinfo {author} {\bibfnamefont {M.}~\bibnamefont {Frasca}},
  \bibinfo {author} {\bibfnamefont {L.~V.}\ \bibnamefont {Gambuzza}}, \ and\
  \bibinfo {author} {\bibfnamefont {P.}~\bibnamefont {H{\"o}vel}},\ }\bibfield
  {title} {\enquote {\bibinfo {title} {Chimera states in time-varying complex
  networks},}\ }\href@noop {} {\bibfield  {journal} {\bibinfo  {journal} {Phys.
  Rev. E}\ }\textbf {\bibinfo {volume} {91}},\ \bibinfo {pages} {022817}
  (\bibinfo {year} {2015})}\BibitemShut {NoStop}%
\bibitem [{\citenamefont {Assenza}\ \emph {et~al.}(2011)\citenamefont
  {Assenza}, \citenamefont {Guti{\'e}rrez}, \citenamefont {G{\'o}mez-Gardenes},
  \citenamefont {Latora},\ and\ \citenamefont {Boccaletti}}]{Assenza2011}%
  \BibitemOpen
  \bibfield  {author} {\bibinfo {author} {\bibfnamefont {S.}~\bibnamefont
  {Assenza}}, \bibinfo {author} {\bibfnamefont {R.}~\bibnamefont
  {Guti{\'e}rrez}}, \bibinfo {author} {\bibfnamefont {J.}~\bibnamefont
  {G{\'o}mez-Gardenes}}, \bibinfo {author} {\bibfnamefont {V.}~\bibnamefont
  {Latora}}, \ and\ \bibinfo {author} {\bibfnamefont {S.}~\bibnamefont
  {Boccaletti}},\ }\bibfield  {title} {\enquote {\bibinfo {title} {Emergence of
  structural patterns out of synchronization in networks with competitive
  interactions},}\ }\href@noop {} {\bibfield  {journal} {\bibinfo  {journal}
  {Sci. Rep.}\ }\textbf {\bibinfo {volume} {1}},\ \bibinfo {pages} {99}
  (\bibinfo {year} {2011})}\BibitemShut {NoStop}%
\bibitem [{\citenamefont {Ghosh}\ \emph {et~al.}(2022)\citenamefont {Ghosh},
  \citenamefont {Frasca}, \citenamefont {Rizzo}, \citenamefont {Majhi},
  \citenamefont {Rakshit}, \citenamefont {Alfaro-Bittner},\ and\ \citenamefont
  {Boccaletti}}]{Ghosh2022}%
  \BibitemOpen
  \bibfield  {author} {\bibinfo {author} {\bibfnamefont {D.}~\bibnamefont
  {Ghosh}}, \bibinfo {author} {\bibfnamefont {M.}~\bibnamefont {Frasca}},
  \bibinfo {author} {\bibfnamefont {A.}~\bibnamefont {Rizzo}}, \bibinfo
  {author} {\bibfnamefont {S.}~\bibnamefont {Majhi}}, \bibinfo {author}
  {\bibfnamefont {S.}~\bibnamefont {Rakshit}}, \bibinfo {author} {\bibfnamefont
  {K.}~\bibnamefont {Alfaro-Bittner}}, \ and\ \bibinfo {author} {\bibfnamefont
  {S.}~\bibnamefont {Boccaletti}},\ }\bibfield  {title} {\enquote {\bibinfo
  {title} {The synchronized dynamics of time-varying networks},}\ }\href@noop
  {} {\bibfield  {journal} {\bibinfo  {journal} {Phys. Rep.}\ }\textbf
  {\bibinfo {volume} {949}},\ \bibinfo {pages} {1--63} (\bibinfo {year}
  {2022})}\BibitemShut {NoStop}%
\bibitem [{\citenamefont {Rothkegel}\ and\ \citenamefont
  {Lehnertz}(2011)}]{rothkegel2011}%
  \BibitemOpen
  \bibfield  {author} {\bibinfo {author} {\bibfnamefont {A.}~\bibnamefont
  {Rothkegel}}\ and\ \bibinfo {author} {\bibfnamefont {K.}~\bibnamefont
  {Lehnertz}},\ }\bibfield  {title} {\enquote {\bibinfo {title} {Recurrent
  events of synchrony in complex networks of pulse-coupled oscillators},}\
  }\href {\doibase 10.1209/0295-5075/95/38001} {\bibfield  {journal} {\bibinfo
  {journal} {Europhys. Lett.}\ }\textbf {\bibinfo {volume} {95}},\ \bibinfo
  {pages} {38001} (\bibinfo {year} {2011})}\BibitemShut {NoStop}%
\bibitem [{\citenamefont {Rothkegel}\ and\ \citenamefont
  {Lehnertz}(2014)}]{rothkegel2014}%
  \BibitemOpen
  \bibfield  {author} {\bibinfo {author} {\bibfnamefont {A.}~\bibnamefont
  {Rothkegel}}\ and\ \bibinfo {author} {\bibfnamefont {K.}~\bibnamefont
  {Lehnertz}},\ }\bibfield  {title} {\enquote {\bibinfo {title} {Irregular
  macroscopic dynamics due to chimera states in small-world networks of
  pulse-coupled oscillators},}\ }\href {\doibase 10.1088/1367-2630/16/5/055006}
  {\bibfield  {journal} {\bibinfo  {journal} {New J. Phys.}\ }\textbf {\bibinfo
  {volume} {16}},\ \bibinfo {pages} {055006} (\bibinfo {year}
  {2014})}\BibitemShut {NoStop}%
\bibitem [{\citenamefont {Gerster}\ \emph {et~al.}(2020)\citenamefont
  {Gerster}, \citenamefont {Berner}, \citenamefont {Sawicki}, \citenamefont
  {Zakharova}, \citenamefont {{\v{S}}koch}, \citenamefont {Hlinka},
  \citenamefont {Lehnertz},\ and\ \citenamefont {Sch{\"o}ll}}]{Gerster2020}%
  \BibitemOpen
  \bibfield  {author} {\bibinfo {author} {\bibfnamefont {M.}~\bibnamefont
  {Gerster}}, \bibinfo {author} {\bibfnamefont {R.}~\bibnamefont {Berner}},
  \bibinfo {author} {\bibfnamefont {J.}~\bibnamefont {Sawicki}}, \bibinfo
  {author} {\bibfnamefont {A.}~\bibnamefont {Zakharova}}, \bibinfo {author}
  {\bibfnamefont {A.}~\bibnamefont {{\v{S}}koch}}, \bibinfo {author}
  {\bibfnamefont {J.}~\bibnamefont {Hlinka}}, \bibinfo {author} {\bibfnamefont
  {K.}~\bibnamefont {Lehnertz}}, \ and\ \bibinfo {author} {\bibfnamefont
  {E.}~\bibnamefont {Sch{\"o}ll}},\ }\bibfield  {title} {\enquote {\bibinfo
  {title} {{FitzHugh--Nagumo} oscillators on complex networks mimic
  epileptic-seizure-related synchronization phenomena},}\ }\href@noop {}
  {\bibfield  {journal} {\bibinfo  {journal} {Chaos: An Interdisciplinary
  Journal of Nonlinear Science}\ }\textbf {\bibinfo {volume} {30}},\ \bibinfo
  {pages} {123130} (\bibinfo {year} {2020})}\BibitemShut {NoStop}%
\bibitem [{\citenamefont {Boaretto}\ \emph {et~al.}(2021)\citenamefont
  {Boaretto}, \citenamefont {Budzinski}, \citenamefont {Rossi}, \citenamefont
  {Manchein}, \citenamefont {Prado}, \citenamefont {Feudel},\ and\
  \citenamefont {Lopes}}]{Boaretto2021}%
  \BibitemOpen
  \bibfield  {author} {\bibinfo {author} {\bibfnamefont {B.}~\bibnamefont
  {Boaretto}}, \bibinfo {author} {\bibfnamefont {R.}~\bibnamefont {Budzinski}},
  \bibinfo {author} {\bibfnamefont {K.}~\bibnamefont {Rossi}}, \bibinfo
  {author} {\bibfnamefont {C.}~\bibnamefont {Manchein}}, \bibinfo {author}
  {\bibfnamefont {T.}~\bibnamefont {Prado}}, \bibinfo {author} {\bibfnamefont
  {U.}~\bibnamefont {Feudel}}, \ and\ \bibinfo {author} {\bibfnamefont
  {S.}~\bibnamefont {Lopes}},\ }\bibfield  {title} {\enquote {\bibinfo {title}
  {Bistability in the synchronization of identical neurons},}\ }\href@noop {}
  {\bibfield  {journal} {\bibinfo  {journal} {Phys. Rev. E}\ }\textbf {\bibinfo
  {volume} {104}},\ \bibinfo {pages} {024204} (\bibinfo {year}
  {2021})}\BibitemShut {NoStop}%
\bibitem [{\citenamefont {Lee}\ and\ \citenamefont
  {Krischer}(2021)}]{LeeKrischner2021}%
  \BibitemOpen
  \bibfield  {author} {\bibinfo {author} {\bibfnamefont {S.}~\bibnamefont
  {Lee}}\ and\ \bibinfo {author} {\bibfnamefont {K.}~\bibnamefont {Krischer}},\
  }\bibfield  {title} {\enquote {\bibinfo {title} {Attracting {P}oisson
  chimeras in two-population networks},}\ }\href@noop {} {\bibfield  {journal}
  {\bibinfo  {journal} {Chaos: An Interdisciplinary Journal of Nonlinear
  Science}\ }\textbf {\bibinfo {volume} {31}},\ \bibinfo {pages} {113101}
  (\bibinfo {year} {2021})}\BibitemShut {NoStop}%
\bibitem [{\citenamefont {Clusella}, \citenamefont {Pietras},\ and\
  \citenamefont {Montbri{\'o}}(2022)}]{Clusella2022}%
  \BibitemOpen
  \bibfield  {author} {\bibinfo {author} {\bibfnamefont {P.}~\bibnamefont
  {Clusella}}, \bibinfo {author} {\bibfnamefont {B.}~\bibnamefont {Pietras}}, \
  and\ \bibinfo {author} {\bibfnamefont {E.}~\bibnamefont {Montbri{\'o}}},\
  }\bibfield  {title} {\enquote {\bibinfo {title} {Kuramoto model for
  populations of quadratic integrate-and-fire neurons with chemical and
  electrical coupling},}\ }\href@noop {} {\bibfield  {journal} {\bibinfo
  {journal} {Chaos: An Interdisciplinary Journal of Nonlinear Science}\
  }\textbf {\bibinfo {volume} {32}},\ \bibinfo {pages} {013105} (\bibinfo
  {year} {2022})}\BibitemShut {NoStop}%
\bibitem [{\citenamefont {Sawicki}\ \emph {et~al.}(2022)\citenamefont
  {Sawicki}, \citenamefont {Hartmann}, \citenamefont {Bader},\ and\
  \citenamefont {Sch{\"o}ll}}]{Sawicki2022}%
  \BibitemOpen
  \bibfield  {author} {\bibinfo {author} {\bibfnamefont {J.}~\bibnamefont
  {Sawicki}}, \bibinfo {author} {\bibfnamefont {L.}~\bibnamefont {Hartmann}},
  \bibinfo {author} {\bibfnamefont {R.}~\bibnamefont {Bader}}, \ and\ \bibinfo
  {author} {\bibfnamefont {E.}~\bibnamefont {Sch{\"o}ll}},\ }\bibfield  {title}
  {\enquote {\bibinfo {title} {Modelling the perception of music in brain
  network dynamics},}\ }\href@noop {} {\bibfield  {journal} {\bibinfo
  {journal} {Front. Netw. Physiol.}\ }\textbf {\bibinfo {volume} {2}},\
  \bibinfo {pages} {910920} (\bibinfo {year} {2022})}\BibitemShut {NoStop}%
\bibitem [{\citenamefont {Thiele}\ \emph {et~al.}(2023)\citenamefont {Thiele},
  \citenamefont {Berner}, \citenamefont {Tass}, \citenamefont {Sch{\"o}ll},\
  and\ \citenamefont {Yanchuk}}]{Thiele2023}%
  \BibitemOpen
  \bibfield  {author} {\bibinfo {author} {\bibfnamefont {M.}~\bibnamefont
  {Thiele}}, \bibinfo {author} {\bibfnamefont {R.}~\bibnamefont {Berner}},
  \bibinfo {author} {\bibfnamefont {P.~A.}\ \bibnamefont {Tass}}, \bibinfo
  {author} {\bibfnamefont {E.}~\bibnamefont {Sch{\"o}ll}}, \ and\ \bibinfo
  {author} {\bibfnamefont {S.}~\bibnamefont {Yanchuk}},\ }\bibfield  {title}
  {\enquote {\bibinfo {title} {Asymmetric adaptivity induces recurrent
  synchronization in complex networks},}\ }\href@noop {} {\bibfield  {journal}
  {\bibinfo  {journal} {Chaos: An Interdisciplinary Journal of Nonlinear
  Science}\ }\textbf {\bibinfo {volume} {33}},\ \bibinfo {pages} {023123}
  (\bibinfo {year} {2023})}\BibitemShut {NoStop}%
\bibitem [{\citenamefont {H{\"o}rsch}\ \emph {et~al.}(2018)\citenamefont
  {H{\"o}rsch}, \citenamefont {Hofmann}, \citenamefont {Schlachtberger},\ and\
  \citenamefont {Brown}}]{Hoersch2018}%
  \BibitemOpen
  \bibfield  {author} {\bibinfo {author} {\bibfnamefont {J.}~\bibnamefont
  {H{\"o}rsch}}, \bibinfo {author} {\bibfnamefont {F.}~\bibnamefont {Hofmann}},
  \bibinfo {author} {\bibfnamefont {D.}~\bibnamefont {Schlachtberger}}, \ and\
  \bibinfo {author} {\bibfnamefont {T.}~\bibnamefont {Brown}},\ }\bibfield
  {title} {\enquote {\bibinfo {title} {{PyPSA-Eur}: {A}n open optimisation
  model of the {European} transmission system},}\ }\href@noop {} {\bibfield
  {journal} {\bibinfo  {journal} {Energy Strategy Rev.}\ }\textbf {\bibinfo
  {volume} {22}},\ \bibinfo {pages} {207--215} (\bibinfo {year}
  {2018})}\BibitemShut {NoStop}%
\bibitem [{\citenamefont {H\"orsch}\ \emph {et~al.}(2022)\citenamefont
  {H\"orsch}, \citenamefont {Neumann}, \citenamefont {Hofmann}, \citenamefont
  {Schlachtberger}, \citenamefont {Frysztacki}, \citenamefont {Hampp},
  \citenamefont {Glaum},\ and\ \citenamefont {Brown}}]{PyPSA-eur-data}%
  \BibitemOpen
  \bibfield  {author} {\bibinfo {author} {\bibfnamefont {J.}~\bibnamefont
  {H\"orsch}}, \bibinfo {author} {\bibfnamefont {F.}~\bibnamefont {Neumann}},
  \bibinfo {author} {\bibfnamefont {F.}~\bibnamefont {Hofmann}}, \bibinfo
  {author} {\bibfnamefont {D.}~\bibnamefont {Schlachtberger}}, \bibinfo
  {author} {\bibfnamefont {M.}~\bibnamefont {Frysztacki}}, \bibinfo {author}
  {\bibfnamefont {J.}~\bibnamefont {Hampp}}, \bibinfo {author} {\bibfnamefont
  {P.}~\bibnamefont {Glaum}}, \ and\ \bibinfo {author} {\bibfnamefont
  {T.}~\bibnamefont {Brown}},\ }\href {\doibase 10.5281/zenodo.7646728}
  {\enquote {\bibinfo {title} {{PyPSA-Eur: An Open Optimisation Model of the
  European Transmission System (Dataset)}},}\ } (\bibinfo {year}
  {2022})\BibitemShut {NoStop}%
\bibitem [{\citenamefont {Christie}(1999)}]{Christie1999}%
  \BibitemOpen
  \bibfield  {author} {\bibinfo {author} {\bibfnamefont {R.~D.}\ \bibnamefont
  {Christie}},\ }\href {http://www.ee.washington.edu/research/pstca/} {\enquote
  {\bibinfo {title} {www.ee.washington.edu/research/pstca/},}\ } (\bibinfo
  {year} {1999})\BibitemShut {NoStop}%
\bibitem [{\citenamefont {Kuramoto}(1975)}]{kuramoto1975}%
  \BibitemOpen
  \bibfield  {author} {\bibinfo {author} {\bibfnamefont {Y.}~\bibnamefont
  {Kuramoto}},\ }\bibfield  {title} {\enquote {\bibinfo {title}
  {Self-entrainment of a population of coupled nonlinear oscillators},}\ }in\
  \href@noop {} {\emph {\bibinfo {booktitle} {International Symposium on
  Mathematical Problems in Theoretical Physics}}},\ \bibinfo {series} {Springer
  Lecture Notes in Physics}, Vol.~\bibinfo {volume} {39},\ \bibinfo {editor}
  {edited by\ \bibinfo {editor} {\bibfnamefont {H.}~\bibnamefont {Araki}}}\
  (\bibinfo {address} {Springer, New York},\ \bibinfo {year} {1975})\ pp.\
  \bibinfo {pages} {420--422}\BibitemShut {NoStop}%
\bibitem [{\citenamefont {Guo}\ \emph {et~al.}(2021)\citenamefont {Guo},
  \citenamefont {Zhang}, \citenamefont {Li}, \citenamefont {Wang},\ and\
  \citenamefont {Yu}}]{Guo2021}%
  \BibitemOpen
  \bibfield  {author} {\bibinfo {author} {\bibfnamefont {Y.}~\bibnamefont
  {Guo}}, \bibinfo {author} {\bibfnamefont {D.}~\bibnamefont {Zhang}}, \bibinfo
  {author} {\bibfnamefont {Z.}~\bibnamefont {Li}}, \bibinfo {author}
  {\bibfnamefont {Q.}~\bibnamefont {Wang}}, \ and\ \bibinfo {author}
  {\bibfnamefont {D.}~\bibnamefont {Yu}},\ }\bibfield  {title} {\enquote
  {\bibinfo {title} {Overviews on the applications of the {K}uramoto model in
  modern power system analysis},}\ }\href@noop {} {\bibfield  {journal}
  {\bibinfo  {journal} {Int. J. Electr. Power Energy Syst.}\ }\textbf {\bibinfo
  {volume} {129}},\ \bibinfo {pages} {106804} (\bibinfo {year}
  {2021})}\BibitemShut {NoStop}%
\bibitem [{\citenamefont {Anvari}\ \emph {et~al.}(2020)\citenamefont {Anvari},
  \citenamefont {Gorj{\~a}o}, \citenamefont {Timme}, \citenamefont {Witthaut},
  \citenamefont {Sch{\"a}fer},\ and\ \citenamefont {Kantz}}]{Anvari2020}%
  \BibitemOpen
  \bibfield  {author} {\bibinfo {author} {\bibfnamefont {M.}~\bibnamefont
  {Anvari}}, \bibinfo {author} {\bibfnamefont {L.~R.}\ \bibnamefont
  {Gorj{\~a}o}}, \bibinfo {author} {\bibfnamefont {M.}~\bibnamefont {Timme}},
  \bibinfo {author} {\bibfnamefont {D.}~\bibnamefont {Witthaut}}, \bibinfo
  {author} {\bibfnamefont {B.}~\bibnamefont {Sch{\"a}fer}}, \ and\ \bibinfo
  {author} {\bibfnamefont {H.}~\bibnamefont {Kantz}},\ }\bibfield  {title}
  {\enquote {\bibinfo {title} {Stochastic properties of the frequency dynamics
  in real and synthetic power grids},}\ }\href@noop {} {\bibfield  {journal}
  {\bibinfo  {journal} {Phys. Rev. Res.}\ }\textbf {\bibinfo {volume} {2}},\
  \bibinfo {pages} {013339} (\bibinfo {year} {2020})}\BibitemShut {NoStop}%
\bibitem [{\citenamefont {Rydin~Gorj{\~a}o}\ \emph {et~al.}(2020)\citenamefont
  {Rydin~Gorj{\~a}o}, \citenamefont {Jumar}, \citenamefont {Maass},
  \citenamefont {Hagenmeyer}, \citenamefont {Yalcin}, \citenamefont {Kruse},
  \citenamefont {Timme}, \citenamefont {Beck}, \citenamefont {Witthaut},\ and\
  \citenamefont {Sch{\"a}fer}}]{RydinGorjao2020}%
  \BibitemOpen
  \bibfield  {author} {\bibinfo {author} {\bibfnamefont {L.}~\bibnamefont
  {Rydin~Gorj{\~a}o}}, \bibinfo {author} {\bibfnamefont {R.}~\bibnamefont
  {Jumar}}, \bibinfo {author} {\bibfnamefont {H.}~\bibnamefont {Maass}},
  \bibinfo {author} {\bibfnamefont {V.}~\bibnamefont {Hagenmeyer}}, \bibinfo
  {author} {\bibfnamefont {G.~C.}\ \bibnamefont {Yalcin}}, \bibinfo {author}
  {\bibfnamefont {J.}~\bibnamefont {Kruse}}, \bibinfo {author} {\bibfnamefont
  {M.}~\bibnamefont {Timme}}, \bibinfo {author} {\bibfnamefont
  {C.}~\bibnamefont {Beck}}, \bibinfo {author} {\bibfnamefont {D.}~\bibnamefont
  {Witthaut}}, \ and\ \bibinfo {author} {\bibfnamefont {B.}~\bibnamefont
  {Sch{\"a}fer}},\ }\bibfield  {title} {\enquote {\bibinfo {title} {Open
  database analysis of scaling and spatio-temporal properties of power grid
  frequencies},}\ }\href@noop {} {\bibfield  {journal} {\bibinfo  {journal}
  {Nature Commun.}\ }\textbf {\bibinfo {volume} {11}},\ \bibinfo {pages} {6362}
  (\bibinfo {year} {2020})}\BibitemShut {NoStop}%
\bibitem [{\citenamefont {Sch{\"a}fer}\ \emph {et~al.}(2023)\citenamefont
  {Sch{\"a}fer}, \citenamefont {Gorj{\~a}o}, \citenamefont {Yalcin},
  \citenamefont {F{\"o}rstner}, \citenamefont {Jumar}, \citenamefont {Maass},
  \citenamefont {K{\"u}hnapfel},\ and\ \citenamefont
  {Hagenmeyer}}]{Schaefer2023}%
  \BibitemOpen
  \bibfield  {author} {\bibinfo {author} {\bibfnamefont {B.}~\bibnamefont
  {Sch{\"a}fer}}, \bibinfo {author} {\bibfnamefont {L.~R.}\ \bibnamefont
  {Gorj{\~a}o}}, \bibinfo {author} {\bibfnamefont {G.~C.}\ \bibnamefont
  {Yalcin}}, \bibinfo {author} {\bibfnamefont {E.}~\bibnamefont
  {F{\"o}rstner}}, \bibinfo {author} {\bibfnamefont {R.}~\bibnamefont {Jumar}},
  \bibinfo {author} {\bibfnamefont {H.}~\bibnamefont {Maass}}, \bibinfo
  {author} {\bibfnamefont {U.}~\bibnamefont {K{\"u}hnapfel}}, \ and\ \bibinfo
  {author} {\bibfnamefont {V.}~\bibnamefont {Hagenmeyer}},\ }\bibfield  {title}
  {\enquote {\bibinfo {title} {Microscopic fluctuations in power-grid frequency
  recordings at the sub-second scale},}\ }\href {\doibase 10.1155/2023/2657039}
  {\bibfield  {journal} {\bibinfo  {journal} {Complexity}\ }\textbf {\bibinfo
  {volume} {2023}},\ \bibinfo {pages} {2657039} (\bibinfo {year}
  {2023})}\BibitemShut {NoStop}%
\bibitem [{\citenamefont {Ansmann}(2018)}]{ansmann2018}%
  \BibitemOpen
  \bibfield  {author} {\bibinfo {author} {\bibfnamefont {G.}~\bibnamefont
  {Ansmann}},\ }\bibfield  {title} {\enquote {\bibinfo {title} {Efficiently and
  easily integrating differential equations with {JiTCODE}, {JiTCDDE}, and
  {JiTCSDE}},}\ }\href@noop {} {\bibfield  {journal} {\bibinfo  {journal}
  {Chaos: An Interdisciplinary Journal of Nonlinear Science}\ }\textbf
  {\bibinfo {volume} {28}},\ \bibinfo {pages} {043116} (\bibinfo {year}
  {2018})}\BibitemShut {NoStop}%
\bibitem [{\citenamefont {Kutil}(2012)}]{Kutil2012}%
  \BibitemOpen
  \bibfield  {author} {\bibinfo {author} {\bibfnamefont {R.}~\bibnamefont
  {Kutil}},\ }\bibfield  {title} {\enquote {\bibinfo {title} {Biased and
  unbiased estimation of the circular mean resultant length and its
  variance},}\ }\href@noop {} {\bibfield  {journal} {\bibinfo  {journal}
  {Statistics}\ }\textbf {\bibinfo {volume} {46}},\ \bibinfo {pages} {549--561}
  (\bibinfo {year} {2012})}\BibitemShut {NoStop}%
\bibitem [{\citenamefont {Pagani}\ and\ \citenamefont
  {Aiello}(2013)}]{pagani2013}%
  \BibitemOpen
  \bibfield  {author} {\bibinfo {author} {\bibfnamefont {G.~A.}\ \bibnamefont
  {Pagani}}\ and\ \bibinfo {author} {\bibfnamefont {M.}~\bibnamefont
  {Aiello}},\ }\bibfield  {title} {\enquote {\bibinfo {title} {The power grid
  as a complex network: a survey},}\ }\href@noop {} {\bibfield  {journal}
  {\bibinfo  {journal} {Physica A: Statistical Mechanics and its Applications}\
  }\textbf {\bibinfo {volume} {392}},\ \bibinfo {pages} {2688--2700} (\bibinfo
  {year} {2013})}\BibitemShut {NoStop}%
\bibitem [{\citenamefont {Amani}\ and\ \citenamefont
  {Jalili}(2021)}]{Amani2021}%
  \BibitemOpen
  \bibfield  {author} {\bibinfo {author} {\bibfnamefont {A.~M.}\ \bibnamefont
  {Amani}}\ and\ \bibinfo {author} {\bibfnamefont {M.}~\bibnamefont {Jalili}},\
  }\bibfield  {title} {\enquote {\bibinfo {title} {Power grids as complex
  networks: Resilience and reliability analysis},}\ }\href@noop {} {\bibfield
  {journal} {\bibinfo  {journal} {IEEE Access}\ }\textbf {\bibinfo {volume}
  {9}},\ \bibinfo {pages} {119010--119031} (\bibinfo {year}
  {2021})}\BibitemShut {NoStop}%
\bibitem [{\citenamefont {{\'O}dor}\ \emph {et~al.}(2022)\citenamefont
  {{\'O}dor}, \citenamefont {Deng}, \citenamefont {Hartmann},\ and\
  \citenamefont {Kelling}}]{Odor2022}%
  \BibitemOpen
  \bibfield  {author} {\bibinfo {author} {\bibfnamefont {G.}~\bibnamefont
  {{\'O}dor}}, \bibinfo {author} {\bibfnamefont {S.}~\bibnamefont {Deng}},
  \bibinfo {author} {\bibfnamefont {B.}~\bibnamefont {Hartmann}}, \ and\
  \bibinfo {author} {\bibfnamefont {J.}~\bibnamefont {Kelling}},\ }\bibfield
  {title} {\enquote {\bibinfo {title} {Synchronization dynamics on power grids
  in {Europe} and the {United States}},}\ }\href@noop {} {\bibfield  {journal}
  {\bibinfo  {journal} {Phys. Rev. E}\ }\textbf {\bibinfo {volume} {106}},\
  \bibinfo {pages} {034311} (\bibinfo {year} {2022})}\BibitemShut {NoStop}%
\bibitem [{\citenamefont {Watts}\ and\ \citenamefont
  {Strogatz}(1998)}]{watts1998}%
  \BibitemOpen
  \bibfield  {author} {\bibinfo {author} {\bibfnamefont {D.~J.}\ \bibnamefont
  {Watts}}\ and\ \bibinfo {author} {\bibfnamefont {S.~H.}\ \bibnamefont
  {Strogatz}},\ }\bibfield  {title} {\enquote {\bibinfo {title} {Collective
  dynamics of `small-world' networks},}\ }\href {\doibase 10.1038/30918}
  {\bibfield  {journal} {\bibinfo  {journal} {Nature}\ }\textbf {\bibinfo
  {volume} {393}},\ \bibinfo {pages} {440--442} (\bibinfo {year}
  {1998})}\BibitemShut {NoStop}%
\bibitem [{\citenamefont {Albert}\ and\ \citenamefont
  {Barab\'asi}(2002)}]{albert2002}%
  \BibitemOpen
  \bibfield  {author} {\bibinfo {author} {\bibfnamefont {R.}~\bibnamefont
  {Albert}}\ and\ \bibinfo {author} {\bibfnamefont {A.-L.}\ \bibnamefont
  {Barab\'asi}},\ }\bibfield  {title} {\enquote {\bibinfo {title} {Statistical
  mechanics of complex networks},}\ }\href {\doibase 10.1103/RevModPhys.74.47}
  {\bibfield  {journal} {\bibinfo  {journal} {Rev. Mod. Phys.}\ }\textbf
  {\bibinfo {volume} {74}},\ \bibinfo {pages} {47--97} (\bibinfo {year}
  {2002})}\BibitemShut {NoStop}%
\bibitem [{\citenamefont {Erd\H{o}s}\ and\ \citenamefont
  {R\'enyi}(1959)}]{erdos1959}%
  \BibitemOpen
  \bibfield  {author} {\bibinfo {author} {\bibfnamefont {P.}~\bibnamefont
  {Erd\H{o}s}}\ and\ \bibinfo {author} {\bibfnamefont {A.}~\bibnamefont
  {R\'enyi}},\ }\bibfield  {title} {\enquote {\bibinfo {title} {On random
  graphs {I}},}\ }\href@noop {} {\bibfield  {journal} {\bibinfo  {journal}
  {Publ. Math. Debrecen}\ }\textbf {\bibinfo {volume} {6}},\ \bibinfo {pages}
  {290--297} (\bibinfo {year} {1959})}\BibitemShut {NoStop}%
\bibitem [{\citenamefont {Johnsonbaugh}\ and\ \citenamefont
  {Kalin}(1991)}]{Johnsonbaugh1991}%
  \BibitemOpen
  \bibfield  {author} {\bibinfo {author} {\bibfnamefont {R.}~\bibnamefont
  {Johnsonbaugh}}\ and\ \bibinfo {author} {\bibfnamefont {M.}~\bibnamefont
  {Kalin}},\ }\bibfield  {title} {\enquote {\bibinfo {title} {A graph
  generation software package},}\ }in\ \href@noop {} {\emph {\bibinfo
  {booktitle} {Proceedings of the twenty-second SIGCSE technical symposium on
  Computer Science Education}}}\ (\bibinfo {year} {1991})\ pp.\ \bibinfo
  {pages} {151--154}\BibitemShut {NoStop}%
\bibitem [{\citenamefont {Boccaletti}\ \emph {et~al.}(2006)\citenamefont
  {Boccaletti}, \citenamefont {Latora}, \citenamefont {Moreno}, \citenamefont
  {Chavez},\ and\ \citenamefont {Hwang}}]{Boccaletti2006}%
  \BibitemOpen
  \bibfield  {author} {\bibinfo {author} {\bibfnamefont {S.}~\bibnamefont
  {Boccaletti}}, \bibinfo {author} {\bibfnamefont {V.}~\bibnamefont {Latora}},
  \bibinfo {author} {\bibfnamefont {Y.}~\bibnamefont {Moreno}}, \bibinfo
  {author} {\bibfnamefont {M.}~\bibnamefont {Chavez}}, \ and\ \bibinfo {author}
  {\bibfnamefont {D.-U.}\ \bibnamefont {Hwang}},\ }\bibfield  {title} {\enquote
  {\bibinfo {title} {Complex networks: Structure and dynamics},}\ }\href
  {\doibase 10.1016/j.physrep.2005.10.009} {\bibfield  {journal} {\bibinfo
  {journal} {Phys. Rep.}\ }\textbf {\bibinfo {volume} {424}},\ \bibinfo {pages}
  {175--308} (\bibinfo {year} {2006})}\BibitemShut {NoStop}%
\bibitem [{\citenamefont {Newman}(2003)}]{Newman2003}%
  \BibitemOpen
  \bibfield  {author} {\bibinfo {author} {\bibfnamefont {M.~E.~J.}\
  \bibnamefont {Newman}},\ }\bibfield  {title} {\enquote {\bibinfo {title} {The
  structure and function of complex networks},}\ }\href {\doibase
  10.1137/S003614450342480} {\bibfield  {journal} {\bibinfo  {journal} {SIAM
  Rev.}\ }\textbf {\bibinfo {volume} {45}},\ \bibinfo {pages} {167--256}
  (\bibinfo {year} {2003})}\BibitemShut {NoStop}%
\bibitem [{\citenamefont {Motter}, \citenamefont {Zhou},\ and\ \citenamefont
  {Kurths}(2005)}]{motter2005b}%
  \BibitemOpen
  \bibfield  {author} {\bibinfo {author} {\bibfnamefont {A.~E.}\ \bibnamefont
  {Motter}}, \bibinfo {author} {\bibfnamefont {C.}~\bibnamefont {Zhou}}, \ and\
  \bibinfo {author} {\bibfnamefont {J.}~\bibnamefont {Kurths}},\ }\bibfield
  {title} {\enquote {\bibinfo {title} {Network synchronization, diffusion, and
  the paradox of heterogeneity},}\ }\href@noop {} {\bibfield  {journal}
  {\bibinfo  {journal} {Phys. Rev. E}\ }\textbf {\bibinfo {volume} {71}},\
  \bibinfo {pages} {016116} (\bibinfo {year} {2005})}\BibitemShut {NoStop}%
\bibitem [{\citenamefont {di~Bernardo}, \citenamefont {Garofalo},\ and\
  \citenamefont {Sorrentino}(2007)}]{bernado2007}%
  \BibitemOpen
  \bibfield  {author} {\bibinfo {author} {\bibfnamefont {M.}~\bibnamefont
  {di~Bernardo}}, \bibinfo {author} {\bibfnamefont {F.}~\bibnamefont
  {Garofalo}}, \ and\ \bibinfo {author} {\bibfnamefont {F.}~\bibnamefont
  {Sorrentino}},\ }\bibfield  {title} {\enquote {\bibinfo {title} {Effects of
  degree correlation on the synchronization of networks of oscillators},}\
  }\href@noop {} {\bibfield  {journal} {\bibinfo  {journal} {Int. J.
  Bifurcation Chaos Appl. Sci. Eng.}\ }\textbf {\bibinfo {volume} {17}},\
  \bibinfo {pages} {3499--3506} (\bibinfo {year} {2007})}\BibitemShut {NoStop}%
\bibitem [{\citenamefont {Pecora}\ and\ \citenamefont
  {Carroll}(1998)}]{pecora1998}%
  \BibitemOpen
  \bibfield  {author} {\bibinfo {author} {\bibfnamefont {L.~M.}\ \bibnamefont
  {Pecora}}\ and\ \bibinfo {author} {\bibfnamefont {T.~L.}\ \bibnamefont
  {Carroll}},\ }\bibfield  {title} {\enquote {\bibinfo {title} {Master
  stability functions for synchronized coupled systems},}\ }\href@noop {}
  {\bibfield  {journal} {\bibinfo  {journal} {Phys. Rev. Lett.}\ }\textbf
  {\bibinfo {volume} {80}},\ \bibinfo {pages} {2109--2112} (\bibinfo {year}
  {1998})}\BibitemShut {NoStop}%
\bibitem [{\citenamefont {Barahona}\ and\ \citenamefont
  {Pecora}(2002)}]{barahona2002}%
  \BibitemOpen
  \bibfield  {author} {\bibinfo {author} {\bibfnamefont {M.}~\bibnamefont
  {Barahona}}\ and\ \bibinfo {author} {\bibfnamefont {L.~M.}\ \bibnamefont
  {Pecora}},\ }\bibfield  {title} {\enquote {\bibinfo {title} {Synchronization
  in small-world systems},}\ }\href {\doibase 10.1103/PhysRevLett.89.054101}
  {\bibfield  {journal} {\bibinfo  {journal} {Phys. Rev. Lett.}\ }\textbf
  {\bibinfo {volume} {89}},\ \bibinfo {pages} {054101} (\bibinfo {year}
  {2002})}\BibitemShut {NoStop}%
\bibitem [{\citenamefont {Donetti}, \citenamefont {Hurtado},\ and\
  \citenamefont {Munoz}(2005)}]{donetti2005}%
  \BibitemOpen
  \bibfield  {author} {\bibinfo {author} {\bibfnamefont {L.}~\bibnamefont
  {Donetti}}, \bibinfo {author} {\bibfnamefont {P.~I.}\ \bibnamefont
  {Hurtado}}, \ and\ \bibinfo {author} {\bibfnamefont {M.~A.}\ \bibnamefont
  {Munoz}},\ }\bibfield  {title} {\enquote {\bibinfo {title} {Entangled
  networks, synchronization, and optimal network topology},}\ }\href {\doibase
  10.1103/PhysRevLett.95.188701} {\bibfield  {journal} {\bibinfo  {journal}
  {Phys. Rev. Lett.}\ }\textbf {\bibinfo {volume} {95}},\ \bibinfo {eid}
  {188701} (\bibinfo {year} {2005})}\BibitemShut {NoStop}%
\bibitem [{\citenamefont {Atay}, \citenamefont {B\i{}y\i{}ko\u{g}lu},\ and\
  \citenamefont {Jost}(2006)}]{atay2006}%
  \BibitemOpen
  \bibfield  {author} {\bibinfo {author} {\bibfnamefont {F.~M.}\ \bibnamefont
  {Atay}}, \bibinfo {author} {\bibfnamefont {T.}~\bibnamefont
  {B\i{}y\i{}ko\u{g}lu}}, \ and\ \bibinfo {author} {\bibfnamefont
  {J.}~\bibnamefont {Jost}},\ }\bibfield  {title} {\enquote {\bibinfo {title}
  {Network synchronization: Spectral versus statistical properties},}\ }\href
  {\doibase 10.1016/j.physd.2006.09.018} {\bibfield  {journal} {\bibinfo
  {journal} {Physica~D}\ }\textbf {\bibinfo {volume} {224}},\ \bibinfo {pages}
  {35--41} (\bibinfo {year} {2006})}\BibitemShut {NoStop}%
\bibitem [{\citenamefont {Cotilla-Sanchez}\ \emph {et~al.}(2012)\citenamefont
  {Cotilla-Sanchez}, \citenamefont {Hines}, \citenamefont {Barrows},\ and\
  \citenamefont {Blumsack}}]{CotillaSanchez2012}%
  \BibitemOpen
  \bibfield  {author} {\bibinfo {author} {\bibfnamefont {E.}~\bibnamefont
  {Cotilla-Sanchez}}, \bibinfo {author} {\bibfnamefont {P.~D.}\ \bibnamefont
  {Hines}}, \bibinfo {author} {\bibfnamefont {C.}~\bibnamefont {Barrows}}, \
  and\ \bibinfo {author} {\bibfnamefont {S.}~\bibnamefont {Blumsack}},\
  }\bibfield  {title} {\enquote {\bibinfo {title} {Comparing the topological
  and electrical structure of the {North American} electric power
  infrastructure},}\ }\href@noop {} {\bibfield  {journal} {\bibinfo  {journal}
  {IEEE Syst. J.}\ }\textbf {\bibinfo {volume} {6}},\ \bibinfo {pages}
  {616--626} (\bibinfo {year} {2012})}\BibitemShut {NoStop}%
\bibitem [{\citenamefont {Monfared}, \citenamefont {Jalili},\ and\
  \citenamefont {Alipour}(2014)}]{Monfared2014}%
  \BibitemOpen
  \bibfield  {author} {\bibinfo {author} {\bibfnamefont {M.~A.~S.}\
  \bibnamefont {Monfared}}, \bibinfo {author} {\bibfnamefont {M.}~\bibnamefont
  {Jalili}}, \ and\ \bibinfo {author} {\bibfnamefont {Z.}~\bibnamefont
  {Alipour}},\ }\bibfield  {title} {\enquote {\bibinfo {title} {Topology and
  vulnerability of the {Iranian} power grid},}\ }\href@noop {} {\bibfield
  {journal} {\bibinfo  {journal} {Physica A: Statistical Mechanics and its
  Applications}\ }\textbf {\bibinfo {volume} {406}},\ \bibinfo {pages} {24--33}
  (\bibinfo {year} {2014})}\BibitemShut {NoStop}%
\bibitem [{\citenamefont {Espejo}, \citenamefont {Lumbreras},\ and\
  \citenamefont {Ramos}(2018)}]{Espejo2018}%
  \BibitemOpen
  \bibfield  {author} {\bibinfo {author} {\bibfnamefont {R.}~\bibnamefont
  {Espejo}}, \bibinfo {author} {\bibfnamefont {S.}~\bibnamefont {Lumbreras}}, \
  and\ \bibinfo {author} {\bibfnamefont {A.}~\bibnamefont {Ramos}},\ }\bibfield
   {title} {\enquote {\bibinfo {title} {Analysis of transmission-power-grid
  topology and scalability, the {European} case study},}\ }\href@noop {}
  {\bibfield  {journal} {\bibinfo  {journal} {Physica A: Statistical Mechanics
  and its Applications}\ }\textbf {\bibinfo {volume} {509}},\ \bibinfo {pages}
  {383--395} (\bibinfo {year} {2018})}\BibitemShut {NoStop}%
\bibitem [{\citenamefont {Hartmann}\ and\ \citenamefont
  {Sug{\'a}r}(2021)}]{Hartmann2021}%
  \BibitemOpen
  \bibfield  {author} {\bibinfo {author} {\bibfnamefont {B.}~\bibnamefont
  {Hartmann}}\ and\ \bibinfo {author} {\bibfnamefont {V.}~\bibnamefont
  {Sug{\'a}r}},\ }\bibfield  {title} {\enquote {\bibinfo {title} {Searching for
  small-world and scale-free behaviour in long-term historical data of a
  real-world power grid},}\ }\href@noop {} {\bibfield  {journal} {\bibinfo
  {journal} {Sci. Rep.}\ }\textbf {\bibinfo {volume} {11}},\ \bibinfo {pages}
  {6575} (\bibinfo {year} {2021})}\BibitemShut {NoStop}%
\bibitem [{\citenamefont {Ansmann}(2015)}]{ansmann2015b}%
  \BibitemOpen
  \bibfield  {author} {\bibinfo {author} {\bibfnamefont {G.}~\bibnamefont
  {Ansmann}},\ }\bibfield  {title} {\enquote {\bibinfo {title} {A highly
  specific test for periodicity},}\ }\href {\doibase 10.1063/1.4934968}
  {\bibfield  {journal} {\bibinfo  {journal} {Chaos}\ }\textbf {\bibinfo
  {volume} {25}},\ \bibinfo {pages} {113106} (\bibinfo {year}
  {2015})}\BibitemShut {NoStop}%
\bibitem [{\citenamefont {Wang}, \citenamefont {Suzuki},\ and\ \citenamefont
  {Aihara}(2016)}]{wang2016c}%
  \BibitemOpen
  \bibfield  {author} {\bibinfo {author} {\bibfnamefont {B.}~\bibnamefont
  {Wang}}, \bibinfo {author} {\bibfnamefont {H.}~\bibnamefont {Suzuki}}, \ and\
  \bibinfo {author} {\bibfnamefont {K.}~\bibnamefont {Aihara}},\ }\bibfield
  {title} {\enquote {\bibinfo {title} {Enhancing synchronization stability in a
  multi-area power grid},}\ }\href@noop {} {\bibfield  {journal} {\bibinfo
  {journal} {Sci. Rep.}\ }\textbf {\bibinfo {volume} {6}},\ \bibinfo {pages}
  {26596} (\bibinfo {year} {2016})}\BibitemShut {NoStop}%
\bibitem [{\citenamefont {Tchuisseu}\ \emph {et~al.}(2018)\citenamefont
  {Tchuisseu}, \citenamefont {Gomila}, \citenamefont {Colet}, \citenamefont
  {Witthaut}, \citenamefont {Timme},\ and\ \citenamefont
  {Sch{\"a}fer}}]{tchuisseu2018}%
  \BibitemOpen
  \bibfield  {author} {\bibinfo {author} {\bibfnamefont {E.~B.~T.}\
  \bibnamefont {Tchuisseu}}, \bibinfo {author} {\bibfnamefont {D.}~\bibnamefont
  {Gomila}}, \bibinfo {author} {\bibfnamefont {P.}~\bibnamefont {Colet}},
  \bibinfo {author} {\bibfnamefont {D.}~\bibnamefont {Witthaut}}, \bibinfo
  {author} {\bibfnamefont {M.}~\bibnamefont {Timme}}, \ and\ \bibinfo {author}
  {\bibfnamefont {B.}~\bibnamefont {Sch{\"a}fer}},\ }\bibfield  {title}
  {\enquote {\bibinfo {title} {Curing {B}raess’ paradox by secondary control
  in power grids},}\ }\href@noop {} {\bibfield  {journal} {\bibinfo  {journal}
  {New J. Phys.}\ }\textbf {\bibinfo {volume} {20}},\ \bibinfo {pages} {083005}
  (\bibinfo {year} {2018})}\BibitemShut {NoStop}%
\bibitem [{\citenamefont {Mardia}\ and\ \citenamefont
  {Jupp}(2000)}]{mardia2000}%
  \BibitemOpen
  \bibfield  {author} {\bibinfo {author} {\bibfnamefont {K.~V.}\ \bibnamefont
  {Mardia}}\ and\ \bibinfo {author} {\bibfnamefont {P.~E.}\ \bibnamefont
  {Jupp}},\ }\href@noop {} {\emph {\bibinfo {title} {Directional Statistics}}}\
  (\bibinfo  {publisher} {Wiley},\ \bibinfo {address} {New York},\ \bibinfo
  {year} {2000})\BibitemShut {NoStop}%
\bibitem [{\citenamefont {Gong}\ \emph {et~al.}(2019)\citenamefont {Gong},
  \citenamefont {Zheng}, \citenamefont {Toenjes},\ and\ \citenamefont
  {Pikovsky}}]{gong2019}%
  \BibitemOpen
  \bibfield  {author} {\bibinfo {author} {\bibfnamefont {C.~C.}\ \bibnamefont
  {Gong}}, \bibinfo {author} {\bibfnamefont {C.}~\bibnamefont {Zheng}},
  \bibinfo {author} {\bibfnamefont {R.}~\bibnamefont {Toenjes}}, \ and\
  \bibinfo {author} {\bibfnamefont {A.}~\bibnamefont {Pikovsky}},\ }\bibfield
  {title} {\enquote {\bibinfo {title} {Repulsively coupled {Kuramoto-Sakaguchi}
  phase oscillators ensemble subject to common noise},}\ }\href@noop {}
  {\bibfield  {journal} {\bibinfo  {journal} {Chaos: An Interdisciplinary
  Journal of Nonlinear Science}\ }\textbf {\bibinfo {volume} {29}},\ \bibinfo
  {pages} {033127} (\bibinfo {year} {2019})}\BibitemShut {NoStop}%
\end{thebibliography}

\end{document}